\documentclass[
twoside,
11pt,
]{article}

\usepackage[utf8x]{inputenc}
\usepackage{amsmath,amsfonts,amsthm,amssymb}
\usepackage[table,dvipsnames]{xcolor}
\usepackage[final]{graphicx}
\usepackage{subfigure}

\usepackage{cases}
\usepackage{multirow}
\usepackage{url}
\usepackage{hyperref}
\usepackage{booktabs}
\usepackage{array}
\usepackage{multirow}
\newcolumntype{C}[1]{>{\centering\arraybackslash}m{#1}}

\usepackage{dashrule}
\usepackage[
 capitalise
]{cleveref}
\usepackage{bm}
\usepackage{lmodern}
\newtheorem{defn}{Definition}[section]

\usepackage{xcolor}
\definecolor{myGreen}{rgb}{0,0.8,0}
\definecolor{myMagenta}{rgb}{1,0,1}

\newcommand{\diag}{\mathop{\boldsymbol{\mathrm{diag}}}}

\hypersetup{
	backref=true,            %permet d'ajouter des liens dans...
	pagebackref=true,    %...les bibliographies
	hyperindex=true,      %ajoute des liens dans les index.
	colorlinks=true,        %colorise les liens
	breaklinks=true,       %permet le retour àla ligne dans les liens trop longs
	urlcolor= blue,        %couleur des hyperliens
	linkcolor= blue,        %couleur des liens internes
	citecolor=Green,
	bookmarks=true,       %cree des signets pour Acrobat
	bookmarksopen=false  %si les signets Acrobat sont crees,
}

\usepackage{geometry}

\graphicspath{{Fig_Article_NES_InfluEps/}}

\usepackage[
font={small},
labelfont=bf,
width=1\textwidth,
labelsep=period]{caption}

\usepackage{sectsty}
\sectionfont{
\large
}

\subsectionfont{
\normalsize\normalfont\itshape
}

\subsubsectionfont{
\normalfont
\itshape
}

\usepackage{fancyhdr}
\fancyheadoffset[LE,RO]{0cm}

\usepackage{titling}
\usepackage[auth-lg]{authblk}
\title{Scaling law for the slow flow of an unstable mechanical system coupled to a nonlinear energy sink}

\author{Baptiste Bergeot\thanks{Corresponding author: \texttt{baptiste.bergeot@insa-cvl.fr}}}
\affil{INSA CVL, Univ. Orl\'{e}ans, Univ. Tours, LaM\'{e} EA 7494, F-41034, 3 Rue de la Chocolaterie, CS 23410, 41034 Blois Cedex, France}

\date{\today}
\begin{document}

%\noindent \textit{Journal of Sound and Vibration 503 (2021): 116109}

\maketitle

\begin{abstract}
In this paper one first shows that the slow flow of a mechanical system with one unstable mode coupled to a Nonlinear Energy Sink (NES) can be reduced, in the neighborhood of a fold point of its critical manifold, to a normal form of the dynamic saddle-node bifurcation. This allows us to then obtain a scaling law for the slow flow dynamics and to improve the accuracy of the theoretical prediction of the mitigation limit of the NES previously obtained as part of a zeroth-order approximation. For that purpose, the governing equations of the coupled system are first simplified using a reduced-order model for the primary structure by keeping only its unstable modal coordinates. The slow flow is then derived by means of the complexification-averaging method and, by the presence of a small perturbation parameter related to the mass ratio between the NES and the primary structure, it appears as a fast-slow system. The center manifold theorem is finally used to obtain the reduced form of the slow flow which is solved analytically leading to the scaling law. The latter reveals a nontrivial dependence with respect to the small perturbation parameter of the slow flow dynamics near the fold point, involving the fractional exponents 1/3 and 2/3. Finally, a new theoretical prediction of the mitigation limit is deduced from the scaling law. In the end, the proposed methodology is exemplified and validated numerically using an aeroelastic aircraft wing model coupled to one NES.
\end{abstract}

%\tableofcontents

%-------------------------------------------------------------------------------------------------%
%-------------------------------------------------------------------------------------------------%
% Inroduction
%-------------------------------------------------------------------------------------------------%
%-------------------------------------------------------------------------------------------------%
\section{Introduction}\label{sec:intro}

The nonlinear energy sinks (NESs) are nowadays well-known devices used for passive shock and vibration mitigation of undesired oscillations caused by either external, parametric or self-excitations of a primary structure. An NES is classically defined as a nonlinear dynamical attachment consisting of a light mass (compared to the total mass of the primary structure), an essentially nonlinear spring (most of the time purely cubic) and a viscous linear damper. The operating of the NES is based on the phenomenon of targeted energy transfer (TET) through which a properly designed strongly nonlinear oscillator can be tuned to any frequency in order to perform an irreversible energy transfer from the primary structure towards itself. In their seminal papers~\cite{Gendelman2001,vakakis2001} Gendelman, Vakakis and co-workers explain the TET phenomenon by the interaction between two nonlinear modes of vibrations of the system producing a $1$:$1$ resonance capture. Reviews of these concepts can be found in~\cite{VakatisBook2009} and more recently in a part of \cite{lu2018}.

The use of NES to mitigate limit cycle oscillations (LCOs) stemming from dynamic instabilities has been widely studied in the past. The first work reported in this framework concerns the mitigation of LCOs of the Van der Pol oscillator~\cite{LeeSCHM2006}. By means of a perturbation analysis, Gendelman and Bar \cite{Gendelman2010Phys} predicted the response regimes of the same system. This work has been extended to a Van der Pol-Duffing oscillator coupled to one NES by Domany and Gendelman~\cite{domany2013}. Numerous papers have been dedicated to the problem of mitigation, by means of one or several NESs, of LCOs due to aeroelastic instabilities. The seminal papers are those of Lee et al.~\cite{LeeAIAApI2007,LeeAIAApI2007b,lee2008} and concerned aeroelastic instabilities in aircraft wing. In these works, the problem has been investigated both numerically and experimentally. The same aircraft wing model coupled to one NES has been studied theoretically using the so-called complexification-averaging method~\cite{Manevitch1999} together with a singular perturbation approach~\cite{Gendelman2010SIAM} or a perturbation algorithm based on a mixed multiple scale-harmonic balance method (MSHBM)~\cite{luongo2014}. Vaurigaud et al.~\cite{vaurigaud2011b} investigated a problem of passive nonlinear TET between a two degrees of freedom long span bridge model prone to coupled flutter and an NES. The mitigation, using an NES, of vortex-induced vibrations resulting from the nonlinear interaction of a laminar flow and a rigid circular cylinder has been first studied by Tumkur et al. ~\cite{Tumkur2013} constructing a two-DOF reduced-order model of the system and validating it by means of a comparison with a full order finite-element model. On the same issue, an improved and experimentally validated reduced model has been proposed by Dai et al.~\cite{Dai2017}. The passive control of helicopter ground resonance instabilities by a means of NESs has been theoretically and numerically studied by Bergeot et al. considering an NES attached to the helicopter fuselage~\cite{Bergeot201672} and to the blades~\cite{Bergeot2017JSV}. Chatter control in machine tool vibrations has been studied by Gourc et al.~\cite{Gourc2013} considering a vibro-impact NES and by~Nankali et al.~\cite{nankali2017} considering a purely cubic NES. In \cite{Bergeot2017a}, mitigation of friction-induced vibrations due to mode coupling instabilities in a friction system has been investigated. The use of several NESs to mitigate LCOs has been analytically investigated by Bergeot et Bellizzi. First, a network of parallel NESs coupled to a Van der Pol oscillator has been analyzed in~\cite{Bergeot2018} and  then the prediction of the steady-state regimes of a multi-DOF dynamical system having one unstable mode and coupled to a set of NESs has been performed in~\cite{bergeot2019}. Finally, in a recent paper, the possibility of mitigating simultaneously two unstable modes of a linear multi-DOF primary system using a single NES has been investigated by Bergeot et al.~\cite{BERCNSNS2020}. The asymptotic analysis proposed in the latter reference reveals that the NES attachment can produce several and complex mitigated responses which results from the presence of several stable solutions and from the competition between them.

In general, when an NES is attached to a primary structure, the resulting coupled model is analyzed by introducing a small perturbation parameter related to the mass ratio between the NES and the primary structure. Under the assumption of a $1$:$1$ resonance capture, the system is first averaged over one period corresponding to a natural frequency of the primary structure. The resulting averaged system, called \textit{slow flow}, is then analyzed by means of singular perturbation techniques (multiple scales method~\cite{nayfeh2008perturbation} or geometric singular perturbation theory~\cite{Jones1995}). The first key point of these analytical treatments is the partition of the slow flow into two time scales (one slow and one fast\footnote{Sometimes the terms ``slow" and ``super-slow"~\cite{Gendelman2010Phys,domany2013} are used to leave the term ``fast" for the oscillations on which the average was carried out and therefore to be coherent with term slow flow. In the present work the terms ``slow" and ``fast" are preferred to be in agreement with the vocabulary usually used in the literature on dynamical systems.}). In this representation the time evolution of the slow flow is thus described as a succession of slow and fast epochs which are described independently. The second key point is the definition of the \textit{critical manifold} whose system trajectories approaches during slow epochs. In general, the analysis is carried out within the zeroth-order approximation, i.e. within the limit case in which the perturbation parameter is equal to zero. Consequently, the obtained analytical results depreciate for the largest values of the perturbation parameter by thus limiting their predictive power. Luongo and Zulli~\cite{luongo2014} used a perturbation technique to express the \textit{invariant manifold} as a power series with respect to the perturbation parameter in which the first term is the critical manifold\footnote{The difference between critical and invariant manifolds will be calrified in the manuscript in \cref{sec:scallaw}.}. As stated by the Fenichel theory~\cite{Fenichel1979}, this representation is relevant for normally hyperbolic branch of the critical manifold but fails, at fold points (the junction between slow and fast epochs takes place in the vicinity of these points), to approach the actual trajectory of the system. Indeed, the literature on dynamical systems shows that the trajectory passes through the region near these fold points with a nontrivial scaling behavior with respect to the perturbation parameter that it is not possible to obtain with the classical perturbation methods (see e.g.~\cite{bk:khuehn2015} Sect. 5.4).

By means of a sophisticated analysis, the originality of the present work is to provide an analytical description of the system dynamic behavior which reports on this nontrivial dependence on the small perturbation parameter. This analytical result is hereafter referred as the scaling law of the slow flow, this notion is clarified below within the main body of the present manuscript. The main objective is to obtain an analytical description of the system with a high predictive power compared to what has been proposed in the past in the literature and then to hope in the future of including them in a NES design procedure.

The equations of motion of the system under study are presented in \cref{sec:eom}. The full order model equations of a mechanical system coupled to one NES are derived in~\cref{sec:IniMod}. Assuming a primary structure with only one unstable mode, the modal reduction technique, leading to the reduced-order dimensionless model, is presented in~\cref{sec:red1}. \Cref{sec:resultlit} summarizes and extends some previous analytical results of the literature obtained within the zeroth-order approximation. \Cref{sec:scallaw} expounds the main results of the paper. In \cref{sec:CMR}, the center manifold theorem is used to reduce the dynamics of the slow flow of the system. This reduced form is solved analytically in \cref{sec:ThSol} leading to the scaling law from which a new analytical prediction of the mitigation limit is derived in \cref{sec:NewAnPred}. In \cref{sec:appl}, the proposed methodology is exemplified and validated numerically using an aeroelastic aircraft wing model coupled to one NES. Finally, concluding remarks are formulated in \cref{sec:ccl}.

%-------------------------------------------------------------------------------------------------%
%-------------------------------------------------------------------------------------------------%
% Section
%-------------------------------------------------------------------------------------------------%
%-------------------------------------------------------------------------------------------------%
\section{Equations of the model}
\label{sec:eom}

The dynamical system under study in this paper is presented in the present section. The following developments are similar to those presented in~\cite{bergeot2019}.

%-------------------------------------------------------------------------------------------------%
% Subsection
%-------------------------------------------------------------------------------------------------%
\subsection{The initial model}
\label{sec:IniMod}

The primary structure under consideration in this work is described by the following general set of nonlinear differential equations
\begin{equation}
{\bf \tilde{M}} {\bf \ddot{\tilde{x}}} +{\bf \tilde{C}} {\bf \dot{\tilde{x}}} + {\bf \tilde{K}} {\bf \tilde{x}} +\bf {\bf \tilde{g}^{NL}\left(\tilde{x} \right)}= \bf 0,
\label{eq:systini}
\end{equation}
where ${\bf \tilde{x}} = \left(\tilde{x}_1,\dots,\tilde{x}_N\right)^T$ with $()^T$ the transpose operator, the dot represents time-differentiation and $\bf \tilde{M}$, $\bf \tilde{C}$ and $\bf \tilde{K}$ are the mass matrix, the damping matrix and the stiffness matrix respectively. The nonlinear vector-valued function ${\bf \tilde{g}^{NL}}$ represents the nonlinearity of the primary system. Moreover, cubic nonlinearities are assumed, i.e. each component of ${\bf \tilde{g}^{NL}}$ is a linear combination of monomial terms of order $3$. In these first equations, the tilde symbol is used before the rescaling which will be carried out below (to obtain variables without tilde after rescaling).

System \eqref{eq:systini} can undergo a dynamic instability of the trivial solution through a Hopf bifurcation, i.e. it has a pair of complex conjugate eigenvalues which cross the complex plane imaginary axis. The loss of stability of the trivial solution comes with the production of a periodic solution and the nonlinear function ${\bf \tilde{g}^{NL}}$ allows the existence of Limit Cycle Oscillations (LCOs) on which the system can saturate. For the purpose of mitigating the LCOs, one purely cubic ungrounded NES with mass $\tilde{m}_{h}$, damping coefficient $\tilde{c}_{h}$ and cubic stiffness $\tilde{k}^\text{NL}_{h}$ is used. Taking into account the NES displacement $\tilde{h}(t)$, the equations of the coupled system are
\begin{subequations}
	\label{eq:systprotec0} 
	\begin{align}
	{\bf \tilde{M}} {\bf \ddot{\tilde{x}}} +{\bf \tilde{C}} {\bf \dot{\tilde{x}}} + {\bf \tilde{K}} {\bf \tilde{x}} + 
	{\bf \tilde{g}^{NL}\left(\tilde{x} \right)}+%\nonumber\\
	{\bf \tilde{B}} \Big(
	\tilde{c}_{h} {\left({\bf A \dot{\tilde{x}}}-\dot{\tilde{h}}\right)}+{\tilde{k}^\text{NL}_{h}}{\left({\bf A \tilde{x}}-\tilde{h} \right)^3}\Big)
	&=\bf0\label{eq:systprotec0a} \\
	\tilde{m}_{h}{\ddot{\tilde{h}}} -
	\tilde{c}_{h} {\left({\bf A \dot{\tilde{x}}}-\dot{\tilde{h}}\right)}-{\tilde{k}^\text{NL}_{h}}\left({\bf A \tilde{x}} -\tilde{h} \right)^3
	&= 0\label{eq:systprotec0b},
	\end{align}
\end{subequations}
where ${\bf A}=\left(a_{1}, \dots, a_{N}\right)$ and ${\bf \tilde{B}}=\left(\tilde{b}_{1}, \dots, \tilde{b}_{N}\right)^T$ are the influence coefficient vectors which depend on the position of the NES.

One assumes now that the mass $\tilde{m}_{h}$ and the damping coefficient $\tilde{c}_{h}$ of the NES are small compared to the mass of the primary structure~\eqref{eq:systini} and also that the latter is weakly nonlinear. We therefore introduce a small perturbation dimensionless parameter $\epsilon$ (with $0<\epsilon\ll1$) and the associated rescaled coefficients ${m}_{h}$ and ${c}_{h}$ and vector function ${\bf g^{NL}}$ as $\tilde{m}_{h} = \epsilon m_{h}$, $\tilde{c}_{h} = \epsilon  c_{h}$ and ${\bf \tilde{g}^{NL}}=\epsilon{\bf g^{NL}}$. Then, the variables  $\bf \tilde{x}$ and $\tilde{h}$ are also rescaled through $\epsilon$ as ${\bf x} =\frac{\bf \tilde{x}}{\sqrt{\epsilon}}$ and $h=\frac{\tilde{h}}{\sqrt{\epsilon}}$. Due to the latter rescaling and because cubic nonlinearities are assumed one has ${\bf g^{NL}(\tilde{x})}=\epsilon^{3/2}{\bf g^{NL}(x)}$. Inserting $m_{h}$, $c_{h}$, ${\bf g^{NL}}$, $\bf x$ and $h$ into \eqref{eq:systprotec0}, multiplying \cref{eq:systprotec0a} by $\bf \tilde{M}^{-1}$ and \cref{eq:systprotec0b} by $1/(\epsilon m_h)$ and finally neglecting terms of order strictly higher than 1 in $\epsilon$ lead to
\begin{subequations}
	\label{eq:systprotec4} 
	\begin{align}
	\ddot{{\bf x}}  +  {\bf C} \dot{{\bf x}} +  {\bf K} {\bf x} +
	\epsilon   {\bf  B} {\ddot{h}}
	&=\bf0\label{eq:systprotec4a} \\
	{\ddot{h}} -
	\eta_h{\bf( A \dot{x}}-\dot{h})- {k^\text{NL}_{h}}\left({\bf A} {\bf x}-h \right)^3
	&= 0.\label{eq:systprotec4b}
	\end{align}
\end{subequations}
with $\bf C=\tilde{M}^{-1}\tilde{C}$, $\bf K=\tilde{M}^{-1}\tilde{K}$, 
${\bf B}= m_h {\bf \tilde{M}^{-1}\tilde{B}}$, $\eta_h=c_{h}/m_{h}$ and $k^\text{NL}_{h}=\tilde{k}^\text{NL}_{h}/m_{h}$.

Note that because of the previous assumptions and rescaling, the nonlinearities of the primary structure are neglected from now. Therefore, nonlinear modal interactions in the primary structure such as internal resonances are not taken into account in this work.

%-------------------------------------------------------------------------------------------------%
% Subsection
%-------------------------------------------------------------------------------------------------%
\subsection{Reduction of the dynamics}
\label{sec:red1}

The system \eqref{eq:systprotec4} is now reduced taking into account that the primary system \eqref{eq:systini} undergoes a single instability of the trivial solution. The objective is to obtain a system with two degrees of freedom (one for the unstable mode of the primary structure and one for the NES) on which one will be able to carry out an analysis which would be impossible on a the original system.

To achieve that, it is first convenient to introduce new coordinates ${{\bf v} = {\bf x}  + \epsilon {\bf B}  h}$ and $w= h - {\bf A} {\bf x}$
giving reciprocally
\begin{subequations}
	\label{eq:u&vb}
	\begin{align}
	{\bf x} &= ({\bf I} + \epsilon {\bf B A})^{-1} ({\bf v} - \epsilon {\bf B} w)
	\approx  ({\bf I} - \epsilon {\bf B A}) ({\bf v} - \epsilon {\bf B} w) \label{eq:u&vba}\\
	h &= (1 + \epsilon {\bf A B})^{-1} ({w} + \epsilon {\bf A v})
	\approx (1 - \epsilon {\bf A B}) ({w} + \epsilon {\bf A v})\label{eq:u&vbb}
	\end{align}
\end{subequations}
where ${\bf I}$ is the identity matrix of size $N$.

Using \cref{eq:u&vb}, \cref{eq:systprotec4} is transformed into the following form
\begin{subequations}
	\label{eq:systprotecuv} 
	\begin{align}
	\ddot{{\bf v}}  + {\bf C} \dot{{\bf v}} + {\bf K} {\bf v} -
	\epsilon\Big[
	{\bf C} {\bf B} \left(  {\bf A} \dot{{\bf v}} + \dot{{w}} \right)
	+ {\bf K} {\bf B} \left( {\bf A} {\bf v} +  {w} \right)
	\Big]	&=\bf0\label{eq:systprotecuva} \\
	\ddot{{w}}  + \eta_{h} \dot{{w}}+ k^\text{NL}_{h}w^3 -
	{\bf A C} \dot{{\bf v}} - {\bf A K} {\bf v} &+\nonumber\\
	\epsilon\Big[
	{\bf A B} \eta_h \dot{{w}} + {\bf A C B} \left( {\bf A} \dot{{\bf v}} +  \dot{{w}}\right) +
	{\bf A K B} \left( {\bf A} {\bf v} +  { w} \right) + {\bf A B} k^\text{NL}_{h}w^3\Big]
	&=0.\label{eq:systprotecuvb} 
	\end{align}
\end{subequations}
where only terms of order lower or equal to 1 in $\epsilon$ were kept.

To retain only the essential features of the single instability, the reduction of the dynamics is performed on \cref{eq:systprotecuva}. To achieve this, the latter, is written in state-space form as follows
\begin{equation}
{\bf \dot{y} = D y }+ \epsilon   \left[{\bf D_1} {\bf y} + {\bf D_1} { w} + {\bf D_3} { \dot{w}}\right]
\label{eq:SystY}
\end{equation}
where ${\bf y} = (v_1,\dots,v_N,\dot{v}_1,\dots,\dot{v}_N)^T$,
\begin{align}
\bf D=
\left(
\begin{array}{c|c}
{\bf 0} & {\bf I}_N \\ \hline
\bf -K &\bf -C
\end{array}
\right), \;
{\bf D_1} = 
\left(
\begin{array}{c|c}
{\bf 0} & {\bf 0} \\ \hline
{\bf K B A}  &{\bf C B A} 
\end{array}
\right), \;
{\bf D_2} =
\left(
\begin{array}{c}
{\bf 0} \\ \hline
{\bf K B} 
\end{array}
\right) \; \text{and} \;
{\bf D_3} =
\left(
\begin{array}{c}
{\bf 0} \\ \hline
{\bf C B}
\end{array}
\right).
\label{eq:matA}
\end{align}

Because the matrix $\bf D$, which characterizes the linear dynamics of the primary system, is not symmetric, its diagonalization must be performed using biorthogonality property of the right eigenvectors ${\bf r}_i$ ($i=1,\dots,2N$) and the left eigenvectors ${\bf l}_j$ ($j=1,\dots,2N$) of $\bf D$. When the eigenvectors are properly normalized, the biorthogonality properties consist in ${\bf L}^T \bf D R={\bm \Lambda} $ where ${\bf R}=\left[{\bf r}_1 \ \overline{{\bf r}}_1 \ \dots \ {\bf r}_N \ \overline{{\bf r}}_N\right]$, ${\bf L}=\left[{\bf l}_1 \ \overline{{\bf l}}_1 \  \dots \ {\bf l}_N \ \overline{{\bf l}}_N\right]$ and ${\bm \Lambda} = \diag(\lambda_{1}, \overline{\lambda}_1,\dots,\lambda_{N}, \overline{\lambda}_N)$ are the right and left eigenvector matrices and the eigenvalue diagonal matrix respectively.

The so-called biorthogonal coordinates are then introduced. They are constituted of $N$ pairs of complex conjugates, $q_n$ and $\overline{q}_n$ ($n=1,\dots,N$), and defined by the following relations
\begin{equation}
{\bf y=Rq} \quad \text{and} \quad {{\bf q=L}^T\bf y},
\label{eq:binormtrans}
\end{equation}
where ${\bf q}= (q_1,\overline{q}_1,\dots,q_N,\overline{q}_n)^T$.

Substituting \cref{eq:binormtrans} into \cref{eq:SystY}, the equations of motion take the form of the following system
\begin{equation}
{\bf \dot{q} = {\bm \Lambda} q}+ \epsilon {\bf L}^T \left[{\bf D_1} {\bf R} {\bf q} + {\bf D_2} {w} + {\bf D_3} {\dot{w}}\right],
\label{eq:SystQ}
\end{equation}
and therefore \cref{eq:systprotecuv} is equivalent to 
\begin{subequations}
	\label{eq:systprotecqv} 
	\begin{align}
	{\bf \dot{q} - {\bm \Lambda} q}- \epsilon {\bf L}^T \left[{\bf D_1} {\bf R} {\bf q} + {\bf D_2} {w} + {\bf D_3} {\dot{w}}\right]&=0\label{eq:systprotecqva} \\
	\ddot{{w}}  + \eta_h \dot{{w}}+ k^\text{NL}_h w^3 &-\nonumber \\
	 {\bf A C} ({\bf R}^\text{DL}{\bf q}^\text{U}+{\bf R}^\text{DR}{\bf q}^\text{D})- {\bf A K}   ({\bf R}^\text{UL}{\bf q}^\text{U}+{\bf R}^\text{UR}{\bf q}^\text{D})&+\nonumber \\
	\epsilon\Big[
	{\bf A B}\eta_h \dot{w} + {\bf A C B} \left( {\bf A}  ({\bf R}^\text{DL}{\bf q}^\text{U}+{\bf R}^\text{DR}{\bf q}^\text{D})+ \dot{w}\right) &+ \nonumber \\
	{\bf A K B} \left( {\bf A} ({\bf R}^\text{UL}{\bf q}^\text{U}+{\bf R}^\text{UR}{\bf q}^\text{D})+  {w} \right) + {\bf A B} k^\text{NL}_h w^3\Big]
	&=0,\label{eq:systprotecqvb} 
	\end{align}
\end{subequations}
where the matrix $\bf R$ and the vector $\bf q $ have been split into a $N \times N$-block matrix and a $N \times 1$-block respectively as
 \begin{equation}
{\bf R=
	\begin{bmatrix}
	{\bf R}^\text{UL} & {\bf R}^\text{UR}  \\
	{\bf R}^\text{DL}  & {\bf R}^\text{DR} 
\end{bmatrix}}
\quad
\text{and}
\quad
{\bf q} =(({\bf q}^\text{U})^T,({\bf q}^\text{D})^T)^T
\label{eq:updown1}
\end{equation}
where ``U", ``D", ``L" and ``R" in previous superscripts must be understood as the initial letters of up, down, left and right respectively.

As previously mentioned, one mode of the primary system can become unstable through a Hopf bifurcation. Without loss of generality, the first mode is chosen to be the potential unstable mode. That means that when the chosen physical bifurcation parameter, denoted $\sigma$, crosses the particular parameter value $\sigma_\text{Hopf}$, called \textit{Hopf bifurcation point}, a pair of complex conjugate eigenvalues crosses the complex plane imaginary axis (without loss of generality this is first pair $(\lambda_1,\overline{\lambda}_1)$) whereas all the other eigenvalues $(\lambda_n,\overline{\lambda}_n)$ (with $n=2,\dots,N$) have strictly negative real parts.

Neglecting the coupling term $\epsilon {\bf L}^T \left[{\bf D_1} {\bf R} {\bf q} + {\bf D_2} {w} + {\bf D_3} {\dot{w}}\right]$, \cref{eq:systprotecqva} reduces to the diagonalized system ${\bf \dot{q}}-{\bm \Lambda } \bf q=0$. Therefore, in this case, the variables $q_n$ and  $\overline{q}_n$ ($n=2,\dots,N$), related to the stable modes, tend to zero. When the coupling term (which is of the order of magnitude of $\epsilon$) is taken into account, the variables $q_n$ and  $\overline{q}_n$ do not vanish but their contributions are very small and can be neglected~\cite{Gendelman2010SIAM,bergeot2019,BERCNSNS2020}. Consequently, all terms related to $q_n$ and $\overline{q}_n$ (with $n=2,\dots,N$) are omitted from further consideration and \cref{eq:SystQ} is reduced to
\begin{equation}
\dot{q}_1 = \lambda_1 q_1 + \epsilon {\bf l}_1^T  \Big({\bf D_1}  ({\bf r}_1 q_1 + {\overline{\bf r}_1} \overline{q}_1)+ {\bf D_2} {w} + {\bf D_3} {\dot{w}}\Big).
\label{eq:SystQ2}
\end{equation}

Finally, grouping \cref{eq:systprotecqvb,eq:SystQ2}, splitting $\lambda_1$ into its real and imaginary parts as $\lambda_1=\lambda_\mathcal{R} +j \lambda_\mathcal{I}$ (with $j^2=-1$) and switching the time scale as $t \rightarrow \lambda_\mathcal{I} t$ yield
\begin{subequations}
	\label{eq:systprotecuvRedf} 
	\begin{align}
	\dot{q}_1 - \left(\frac{\lambda_\mathcal{R}}{\lambda_\mathcal{I}} +j\right) q_1 -
	\frac{\epsilon}{\lambda_\mathcal{I} } {\bf l}_1^T  \Big({\bf D_1} ({\bf r}_1 q_1 + \overline{{\bf r}}_1 \overline{q}_1) + {\bf D_2} {w} + {\bf D_3} {\dot{w}}\Big)&=0\\
	\label{eq:systprotecuvRedfb} 
	\ddot{{w}} + \frac{\eta_h}{\lambda_\mathcal{I}  }\dot{{w}}  + \frac{k_h^{NL}}{\lambda_\mathcal{I} ^2}w^3&-\nonumber\\
	\frac{1}{\lambda_\mathcal{I} ^2} \left({\bf A C} ({{\bf r}^\text{DL}_1} q_1 + {\overline{\bf r}^\text{DL}_1} \overline{q}_1) + {\bf A K} ({{\bf r}^\text{UL}_1} q_1 + {\overline{\bf r}^\text{UL}_1} \overline{q}_1)\right)&+ \nonumber \\
	\epsilon\Big[
	{\bf A B}\frac{\eta_h}{\lambda_\mathcal{I}  }\dot{{w}} + \frac{1}{\lambda_\mathcal{I} ^2} {\bf A C B} \left( {\bf A} ({{\bf r}^\text{DL}_1} q_1 + {\overline{\bf r}^\text{DL}_1} \overline{q}_1) +  \lambda_\mathcal{I} \dot{w}\right) &+\nonumber\\
	\frac{1}{\lambda_\mathcal{I} ^2} {\bf A K B} \left( {\bf A} ({{\bf r}^\text{UL}_1} q_1 + {\overline{\bf r}^\text{UL}_1} \overline{q}_1) +  {w} \right) + 
	{\bf A B} \frac{k_h^{NL}}{\lambda_\mathcal{I} ^2} w^3\Big]
	&=0.
	\end{align}
\end{subequations}
where the vectors ${\bf r}^\text{UL}_1$ and ${\bf r}^\text{DL}_1$ correspond to the first columns of the matrices ${\bf R}^\text{DL}$ and ${\bf R}^\text{UL}$ respectively (see \cref{eq:updown1}).

The system of equations \eqref{eq:systprotecuvRedf} is the system under study in the following mathematical developments.

%-------------------------------------------------------------------------------------------------%
%-------------------------------------------------------------------------------------------------%
% Section
%-------------------------------------------------------------------------------------------------%
%-------------------------------------------------------------------------------------------------%
\section{Asymptotic analysis: the zeroth-order approximation of the slow flow}
\label{sec:resultlit}

The present section summarizes and extends previous results of the literature~\cite{Gendelman2010SIAM,Bergeot2018,bergeot2019}. In particular, an analytical expression of the mitigation limit (hereafter called the zeroth-order approximation of the mitigation limit) is derived at the end of the section under several assumptions. The first is given just below.

In \cref{eq:systprotecuvRedf}, the quantities $\lambda_\mathcal{R}$, $\lambda_\mathcal{I}$, the vectors ${\bf l}_1^T$, ${\bf r}_1$, ${\bf r}^\text{UL}_1$ and ${\bf r}^\text{DL}_1$ and the matrices $\bf C$, $\bf K$ (and those which depend on them) depend in general on the bifurcation parameter under consideration~$\sigma$. In this work one assumes that the NES effects are not too far from the Hopf bifurcation point $\sigma_\text{Hopf}$ and therefore the latter are evaluated at $\sigma_\text{Hopf}$ (using the notation $()_\text{Hopf}$) expect $\lambda_\mathcal{R}$ which is by definition equal to zero at $\sigma_\text{Hopf}$. $\lambda_\mathcal{R}$ is evaluated at the actual value of $\sigma$ and the dimensionless eigenvalue real part $\lambda_\mathcal{R}/\omega$ (where $\omega=\lambda_{\mathcal{I},\text{Hopf}}$, the imaginary part of $\lambda_1$ evaluated at $\sigma_\text{Hopf}$) is supposed to be small (i.e. of the order of $\mathcal{O}(\epsilon)$). The parameter $\rho$ is then finally introduced as
\begin{equation}
\rho=\frac{\lambda_\mathcal{R}}{\epsilon\omega} 
\label{eq:rhodef}
\end{equation}
where $\rho$ is of the order of $\mathcal{O}(1)$.

The rescaled dimensionless eigenvalue real part $\rho$ of the unstable mode is considered as a generalized bifurcation parameter (that results from the physical bifurcation parameter $\sigma$ of the initial mechanical system) from which one will follow the changes in the system dynamic behavior.

%------------------------- Subsection -------------------------%
\subsection{Governing equations of the slow flow}
\label{sec:slowflow}

It is know~\cite{Gendelman2001,vakakis2001} that TET appears in the vicinity of the $1$:$1$ resonance between two nonlinear modes of the coupled structure, the latter resulting from the interaction between a linear mode of the primary structure (here this is the considered unstable mode) and the nonlinear mode of the NES. This phenomenon is called $1$:$1$ resonance capture and occurs at a frequency close to the natural frequency of the linear primary structure. To investigate the solution in the neighborhood of this $1$:$1$ resonance capture the complexification-averaging method~\cite{Manevitch1999,VakatisBook2009} is used. The resulting complex averaged dynamics is called slow flow.

The first step consists in introducing the following complex variables
\begin{align}
\label{eq:CAX}
\psi &= {\dot{w}}+ j {w}
\end{align}
or equivalently
\begin{equation}
{w}= \frac{\psi -\overline{\psi}}{2j}, \quad {\dot{w}}= \frac{\psi +\overline{\psi}}{2} \quad \text{and} \quad {\ddot{w}}= \dot{\psi}-\frac{j}{2}\left(\psi +\overline{\psi}\right)
\label{eq:vTOpsi}
\end{equation}
and expressing the complex variables $q_1$ and $\psi$ as
\begin{align}
q_1= \tilde{\phi}_1 e^{j t} \quad \text{and} \quad \psi= \phi_2 e^{j t}.
\label{eq:SlowFastPart}
\end{align}

Substituting \eqref{eq:SlowFastPart} into \eqref{eq:systprotecuvRedf} and averaging over the period $T=2 \pi$ leads to the following system of complex slow modulated amplitudes
\begin{subequations}
\label{eq:slowcomplex}
\begin{align}
\dot{\tilde{\phi}}_1  &=\epsilon\left[
\left(\rho + {\bf l}_{1,\text{Hopf}}^T  {\bf D_{1,\text{Hopf}}} {\bf r}_{1,\text{Hopf}}\right) \tilde{\phi}_1  +\frac{1}{2} \left(-j{\bf l}_{1,\text{Hopf}}^T{\bf D_{2,\text{Hopf}}}  + {\bf l}_{1,\text{Hopf}}^T{\bf D_{3,\text{Hopf}}} \right) \phi_2\right]\\
\dot{\phi}_2 &= \frac{{\bf A K_\text{Hopf}} {{\bf r}^\text{UL}_{1,\text{Hopf}}} + {\bf A C_\text{Hopf}} {{\bf r}^\text{DL}_{1,\text{Hopf}}}}{\omega^2} \tilde{\phi}_1  - \frac{1}{2}\left( j + \mu- j \frac{ 3 \alpha}{4}|\phi_2|^2\right)\phi_2 \nonumber
\\&-\epsilon \Bigg[\frac{1}{\omega^2} \left({\bf A K_\text{Hopf} B T}{{\bf r}^\text{UL}_{1,\text{Hopf}}} + \frac{1}{2} {\bf A C_\text{Hopf} B T} {{\bf r}^\text{DL}_{1,\text{Hopf}}} \right) \tilde{\phi}_1\nonumber\\
&\quad+\frac{1}{2}\left( \frac{\bf A C_\text{Hopf} B}{\omega} - j
\frac{\bf A K_\text{Hopf} B}{\omega^2}   +  {\bf A B} \left(\mu- j \frac{ 3 \alpha}{4}|\phi_2|^2\right)\right) \phi_2
\Bigg]\label{eq:slowcomplexb}
\end{align}
\end{subequations}
with
\begin{equation}
\mu=\frac{\eta_h}{\omega }\quad \text{and} \quad \alpha=\frac{k_h^{NL}}{\omega^2}
\label{eq:parammualp}
\end{equation}
and recalling that the notation $()_\text{Hopf}$ means that the considered quantity is evaluated at the Hopf bifurcation point $\sigma_\text{Hopf}$.

Introducing the variable $\phi_1=2\left({\bf A K_\text{Hopf}} {{\bf r}^\text{UL}_{1,\text{Hopf}}} + {\bf A C_\text{Hopf}} {{\bf r}^\text{DL}_{1,\text{Hopf}}}\right)\tilde{\phi}_1/\omega^2$, and ignoring the terms of the order of magnitude of $\epsilon$ in \eqref{eq:slowcomplexb}\footnote{These terms do not affect qualitatively the dynamic behavior of the slow flow.}, \cref{eq:slowcomplex} reduces to the following synthetic form
\begin{subequations}
\label{eq:slowcomplex2}
\begin{align}
\dot{\phi}_1 +\epsilon\left[\left(a-\rho\right)\phi_1+b \phi_2\right]&=0\\
\dot{\phi}_2+\frac{1}{2}\phi_1 +\frac{1}{2}(\mu+j)\phi_2+j\frac{3}{8}\alpha\phi_2|\phi_2|^2&=0.
\end{align}
\end{subequations}
with the complex coefficients $a$ and $b$ defined as
\begin{equation} 
\begin{split}
a&=-{\bf l}_{1,\text{Hopf}}^T  {\bf D_{1,\text{Hopf}}} {\bf r}_{1,\text{Hopf}}\\
b&=\frac{1}{\omega^2} \left(-j{\bf l}_{1,\text{Hopf}}^T{\bf D_{2,\text{Hopf}}}  + {\bf l}_{1,\text{Hopf}}^T{\bf D_{3,\text{Hopf}}} \right)\left({\bf A K_\text{Hopf}} {{\bf r}^\text{UL}_{1,\text{Hopf}}} + {\bf A C} {{\bf r}^\text{DL}_{1,\text{Hopf}}}\right).
\end{split}
\end{equation}

Then, substituting $\phi_1$ and $\phi_2$ in \cref{eq:slowcomplex} by their polar coordinates defined as $\phi_1 = r e^{j \theta_1}$ and $\phi_2 = s e^{j \theta_2}$, new equations of motion for the real amplitudes $r$ and $s$ and phase difference $\Delta=\theta_2 - \theta_1$ can be obtained as
\begin{subequations}
	\label{eq:slowReal}
	\begin{align}
	\dot{r} &=\epsilon f(r,s,\Delta)\\
	\dot{s} &=\tilde{g}_1(r,s,\Delta,\epsilon)\\
	\dot{\Delta} &=\tilde{g}_2(r,s,\Delta,\epsilon)
	\end{align}
\end{subequations}
with
\begin{subequations}
\begin{align}
f(r,s,\Delta)&=\left(\rho-a_\mathcal{R}\right)r-\left(b_\mathcal{R} \cos \Delta-b_\mathcal{I} \sin \Delta\right)s,\\
\tilde{g}_1(r,s,\Delta,\epsilon)&=-\frac{1}{2}\left(\mu s+r \cos\Delta\right),\\ 
\tilde{g}_2(r,s,\Delta,\epsilon)&=\frac{3}{8} \alpha s^2-\frac{1}{2}+\frac{r}{2 s} \sin\Delta\
+\epsilon \left[a_\mathcal{I}+b_\mathcal{I} \frac{\cos \Delta}{r}+b_\mathcal{R} \frac{\sin \Delta}{r}\right]
\end{align}
\end{subequations}
where the complex parameters $a$ and $b$ were split into real and imaginary parts as $a=a_\mathcal{R}+ja_\mathcal{I}$ and $b=b_\mathcal{R}+jb_\mathcal{I}$.

Because of the presence of the small parameter $\epsilon$, the slow flow is governed by different time scales. More precisely, within the framework of the geometric singular perturbation theory~\cite{Jones1995}, \cref{eq:slowReal} appears as a $(2, 1)$-fast-slow system where $s$ and $\Delta$ are the fast variables and $r$ the slow variable. The latter is related to the modal component $q_1$ whereas the fast variables are associated with the relative displacement $w$ between the NES and the primary structure.

%------------------------- Subsection -------------------------%
\subsection{Critical manifold and fold points of the slow flow}
\label{sec:critman}

The real form of the slow flow, i.e. \cref{eq:slowReal}, is written with respect to the dimensionless slow time $\tau=\epsilon t$ as follows
\begin{subequations}
	\label{eq:slowRealSlow}
	\begin{align}
	r' &= f(r,s,\Delta)\label{eq:slowRealSlowc}\\
	\epsilon s' &=\tilde{g}_1(r,s,\Delta,\epsilon)\label{eq:slowRealSlowa}\\
	\epsilon\Delta' &=\tilde{g}_2(r,s,\Delta,\epsilon)\label{eq:slowRealSlowb}
	\end{align}
\end{subequations}
where $(.)'$ denotes the derivative with respect to the slow time $\tau$. Considering $\epsilon=0$ yields the \textit{slow subsystem}
\begin{subequations}
	\label{eq:slowsubsys}
	\begin{align}
	r' &= f(r,s,\Delta)\label{eq:slowsubsysc}\\
	0&=\tilde{g}_1(r,s,\Delta,0)\label{eq:slowsubsysa}\\
	0 &=\tilde{g}_2(r,s,\Delta,0)\label{eq:slowsubsysb}
	\end{align}
\end{subequations}
which is a differential-algebraic equation, and the \textit{fast subsystem}
\begin{subequations}
\label{eq:fastsubsys}
	\begin{align}
	\dot{r} &=0\\
	\dot{s} &=\tilde{g}_1(r,s,\Delta,0)\\
	\dot{\Delta} &=\tilde{g}_2(r,s,\Delta,0)
	\end{align}
\end{subequations}

The critical manifold of the slow flow is the solution of the algebraic part of \cref{eq:slowsubsys} and it is expressed as follows
\begin{equation}
\label{eq:CM00}
\mathcal{M}_0=\Big\{\left(r,s,\Delta\right) \in \mathbb{R}^{+^2}\times [-\pi , \pi] \; \Big| \;
\tilde{g}_1(r,s,\Delta,0) =0 \;\; \text{and} \;\; \tilde{g}_2(r,s,\Delta,0) =0  \Big\}.
\end{equation}

%\cref{eq:slowsubsysa,eq:slowsubsysb} are written as
%\begin{subequations}
%	\label{eq:slowsubsys2}
%	\begin{align}
%	\tilde{g}_1(r,s,\Delta,0)&=H_1(s)-\frac{1}{2}r\cos\Delta=0\label{eq:slowsubsysa2}\\
%	\tilde{g}_2(r,s,\Delta,0)&=\frac{1}{s}\left(H_2(s)+\frac{1}{2}r\sin\Delta\right)=0\label{eq:slowsubsysb2}
%	\end{align}
%\end{subequations}
%or equivalently as
%\begin{subequations}
%\label{eq:SIMCosSin}
%\begin{align}
%\cos \Delta &=2\frac{H_1(s)}{r} =-\mu \frac{s}{r}\\
%\sin \Delta &=-2\frac{H_2(s)}{r} =\frac{s}{r}\left(1-\frac{3}{4}\alpha s^2\right).
%\end{align}
%\end{subequations}
Combining \cref{eq:slowsubsysa,eq:slowsubsysb} leads to the following modulus and argument equations
\begin{subequations}
\label{eq:RealCM}
\begin{align}
	r& =s \sqrt{\mu ^2+\left(1-\frac{3 \alpha  s^2}{4}\right)^2}= H(s)\label{eq:RealCMa}\\
	\tan\Delta &=\frac{3 \alpha  s^2-4}{4 \mu }.
\end{align}
\end{subequations}

Stability analysis of the critical manifold is now carried out considering the fast subsystem~\eqref{eq:fastsubsys}. It can be shown that a fixed point of~\eqref{eq:fastsubsys} is stable if $d_sH(s)> 0$ and unstable if $d_sH(s) < 0$. Hence, the subset of $\mathcal{M}_0$ satisfying  $d_sH(s) < 0$ defines the attracting zone for the fast dynamics whereas the  subset of $\mathcal{M}_0$ satisfying  $d_sH(s)> 0$ define the repelling zone. 

Exploiting the polynomial properties of $H$, it can be shown that  the local extrema of  $H$ (i.e $d_sH(s)=0$) occur at
\begin{equation}
s^\text{LF} =\frac{2}{3 \sqrt{\alpha}} \sqrt{2- \sqrt{ 1-3\mu^2 }}
\quad \text{and} \quad
s^\text{RF} =\frac{2}{3 \sqrt{\alpha}} \sqrt{2+ \sqrt{ 1-3 \mu^2}}
\label{eq:rNrm}
\end{equation}
if the following relation holds
\begin{equation}
\label{eq:rocond}
\mu < \frac{1}{\sqrt{3}}.
\end{equation}
In \cref{eq:rNrm}, $s^\text{LF}$ and $s^\text{RF}$ are the abscissa values in the $(s,r)$-plane of the a maximum and the minimum of the function $H$ respectively. The superscripts $()^\text{LF}$ and $()^\text{RF}$ refer to \textit{left fold point} and \textit{right fold point} respectively. Indeed, in the $(s,r,\Delta)$-space, the two points $(s^\text{RF},r^\text{RF},\Delta^\text{RF})$ and $(s^\text{LF},r^\text{LF},\Delta^\text{LF})$ (where $r^\text{LF}$, $r^\text{RF}$, $\Delta^\text{LF}$ and $\Delta^\text{RF}$ can be deduced from $s^\text{LF}$ and $s^\text{RF}$ using \cref{eq:RealCM}) are generally called \textit{fold points}.

When the condition~\eqref{eq:rocond} is not satisfied the function $H$ no longer has local extrema. In the remaining of the paper, one considers situations in which \eqref{eq:rocond} holds.

\begin{figure}[t!]%\sidecaption
	\centering
	\includegraphics[width=0.55\columnwidth]{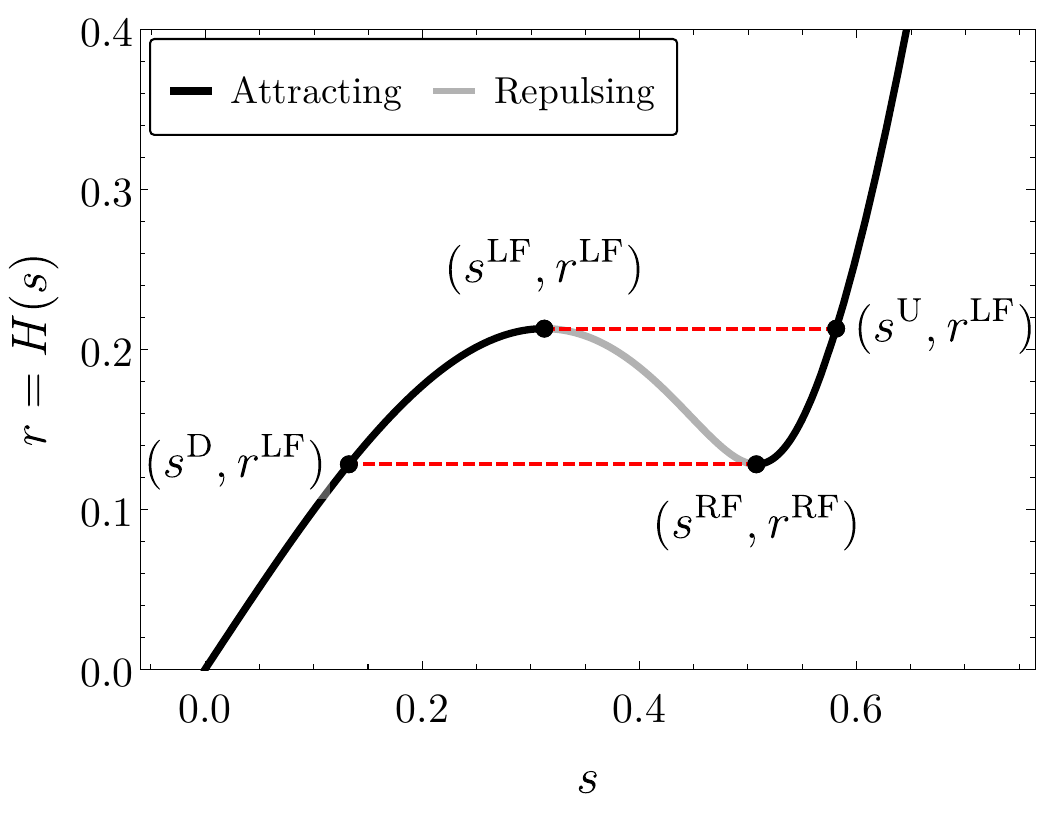}
	\caption{Typical example of the critical manifold in the $(s,r)$-plane given by \cref{eq:RealCMa} for $\mu=0.25$ and $\alpha=5$.}
	\label{fig:CM0}
\end{figure}

A typical critical manifold (more precisely its real part defined by \cref{eq:RealCMa}) is depicted in \cref{fig:CM0}. One can see that the fold points connect attracting parts to repelling part of the critical manifold $\mathcal{M}_0$. The two scalars $s^\text{D}$ and $s^\text{U}$, which are the horizontal projection of the fold points on the critical manifold, are defined by $H\left(s^\text{RF}\right)= H\left(s^\text{D}\right)$ and $H\left(s^\text{LF}\right)= H\left(s^\text{U}\right)$ giving
\begin{equation}
	s^\text{D}= \frac{2\sqrt{2}}{3\sqrt{\alpha}}\sqrt{1-\sqrt{1-3 \mu^2}}
	\quad \text{and} \quad
	s^\text{U}= \frac{2\sqrt{2}}{3\sqrt{\alpha}}\sqrt{1+\sqrt{1-3\mu^2}}.
\label{eq:N2}
\end{equation}

%------------------------- Subsection -------------------------%
\subsection{Fixed points and fold singularities of the slow flow}

Because $0<\epsilon\ll 1$, the fixed points of the slow flow~\eqref{eq:slowReal}, which characterize periodic solutions of \eqref{eq:systprotecuvRedf}, can be approximated by those of the slow subsystem~\eqref{eq:slowsubsys}. From that zeroth-order approximation an analytical expression of these fixed points can be obtained. Note that within this approximation, the fixed points are necessarily located on the critical manifold.

First, substituting \cref{eq:RealCMa} into \cref{eq:slowsubsysc} results in
$\left(H(s)\right)'=f\left(H(s),s,\Delta\right)$
which reduces to
\begin{equation}
\label{eq:SysAsAno12}
d_s H(s)s'=f_s(s),
\end{equation}
where the explicit expression of the function $f_s$ is not given here.
 
Then, a (regular) fixed point of \cref{eq:SysAsAno12} is defined as the roots of $f_{s}\left(s\right)=0$ with $d_s H(s) \neq 0$. Solving $f_s(s)=0$ gives two fixed points (in addition to the trivial solution) whose analytical expressions are
\begin{subequations}\label{eqPFSF}
\begin{align}
s^e_1&=\Bigg(\frac{2}{3\alpha  \left(a_\mathcal{R}-\rho \right)}\Big(2 a_\mathcal{R}-b_\mathcal{I}-2 \rho 
+\sqrt{b_\mathcal{I}^2-4 \mu  \left(a_\mathcal{R}-\rho \right) \left(\mu  a_\mathcal{R}-b_\mathcal{R}-\mu  \rho \right)}\Big)\Bigg)^{1/2},
\label{eqPFSFb}\\
s^e_2&=\Bigg(\frac{2}{3\alpha  \left(a_\mathcal{R}-\rho \right)}\Big(2 a_\mathcal{R}-b_\mathcal{I}-2 \rho 
-\sqrt{b_\mathcal{I}^2-4 \mu  \left(a_\mathcal{R}-\rho \right) \left(\mu  a_\mathcal{R}-b_\mathcal{R}-\mu  \rho \right)}\Big)\Bigg)^{1/2},
\end{align}
\end{subequations}
with $s^e_1<s^e_2$. As usual, the stability of the fixed points are found looking for the sign of $d_s\left(\frac{f_s(s)}{H'(s)}\right)$.

Both fixed points $s^e_1$ and $s^e_2$ no longer exist when they become complex for 
\begin{equation}
\rho^\text{S}=\frac{2 \mu  a_\mathcal{R}+\sqrt{b_\mathcal{I}^2+b_\mathcal{R}^2}-b_\mathcal{R}}{2 \mu }
\label{eq:exPF}
\end{equation}
where the superscript $()^\text{S}$ has no particular meaning here, it is just there to give a name to this particular value of $\rho$.

A singular fixed points of \cref{eq:SysAsAno12}, also called folded singularites, are defined as the roots of the following nonlinear equations $f_{s}\left(s\right)=0$ and $d_s H(s)= 0$. Folded singularities are hints of particular solutions of such fast-slow systems, called canard cycles~\cite{ChasCanard}. These solutions are not investigated in this paper.

%------------------------- Subsection -------------------------%
\subsection{The possible responses}
\label{sec:respregimes}

Previous analysis provides a qualitative description of the slow flow dynamics in the limit case where $\epsilon=0$, i.e. within a zeroth-order approximation. Within this approximation, the slow flow evolves at two time scales: a slow time scale in which the slow flow is on the critical manifold and described by the slow subsystem~\eqref{eq:slowsubsys} and a fast time scale in which the slow flow is outside the critical manifold and described by the fast subsystem~\eqref{eq:fastsubsys}. The particular $S$-shape of the critical manifold together with the stability analysis of the fixed points of the slow flow allows us to explain and predict its different responses and consequently those of the initial full order system~\eqref{eq:systprotecuvRedf}. As has been widely discussed in the literature (see e.g.~\cite{Gendelman2010Phys,Gendelman2010SIAM}), four scenarios are possible. In previous works by the authors~\cite{Bergeot2018,bergeot2019} these responses are classified into two categories. In the first category, the NES acts, resulting in the three responses called \textit{harmless situations}. First, the \textit{Complete suppression} in which the trivial fixed point of the slow flow~\eqref{eq:slowReal} is stable and then reached. In this case, the system is stabilized because of the additional part of the NES including mass and damping. In general, because a light mass and a small damping are considered for the NES, this stabilization effect is negligible, in term of bifurcation parameter range, compared to the mitigation effects described hereinafter. In other words, the Hopf bifurcation point of the primary structure without NES is usually very slightly smaller than the Hopf bifurcation point of the coupled system. Of course, in a context of nonlinear vibrations absorption, the complete suppression is not the desired effect. Then, one can observe \textit{mitigation through periodic response}. In this case, a nontrivial stable fixed point of the slow flow~\eqref{eq:slowReal} is reached, that corresponds to a periodic regime for the initial system~\eqref{eq:systprotecuvRedf}. The last harmless situations correspond to \textit{mitigation through Strongly Modulated Responses (SMRs)}. SMRs correspond to relaxations oscillations of the slow flow~\eqref{eq:slowReal}, that corresponds to a quasi-periodic (amplitude and phase modulated) regime for the initial system~\eqref{eq:systprotecuvRedf}. In the second category, called \textit{harmful situation}, the NES is not able to produce small amplitude responses. For the initial nonlinear system one observes in this case a limit cycle with an amplitude close to that of the system without NES. Note that if the nonlinearity of the primary structure is neglected in the governing equations of the slow flow (this the case in the present work), the harmful situation corresponds for it to an unbounded regime in which its trajectory growths to infinity.

In this paper, we are particularly interested in the transition from SMR to no mitigation which is also, in general, the transition from harmless to harmful situations. Therefore one recalls here briefly what relaxation oscillations of the slow flow are. Note that the following description holds in the limit case for which $\epsilon=0$. In the $(s,r)$-plane, from an initial condition near zero and outside the critical manifold, the trajectory of the slow flow evolves rapidly and horizontally to the left attracting branch of $\mathcal{M}_0$ (this fast epoch is described by the fast subsystem \eqref{eq:fastsubsys}). Then, the slow flow evolves slowly on this branch (this slow epoch is described by the slow subsystem \eqref{eq:slowsubsys}) and, if there is no stable fixed points on it, the trajectory reached the left fold point $(s^\text{LF},r^\text{LF})$ at which $\mathcal{M}_0$ becomes repelling. From the fold point, the slow flow undergoes a fast horizontal jump to $(s^\text{U},r^\text{LF})$ on the right attracting part of $\mathcal{M}_0$ and then a second slow epoch to the right fold point $(s^\text{RF},r^\text{RF})$. A second horizontal jump occurs from $(s^\text{RF},r^\text{RF})$ to $(s^\text{D},r^\text{RF})$ and the trajectory returns to the left attracting part of the $\mathcal{M}_0$. A third slow epoch occurs on $\mathcal{M}_0$ to $(s^\text{LF},r^\text{LF})$ and so on.

%------------------------- Subsection -------------------------%
\subsection{Zeroth-order analytical prediction of the mitigation limit}
\label{sec:predzero}

The mitigation limit is defined below.

\begin{defn}
Considering a set of initial conditions (for the slow flow) as a small perturbation of the trivial solution, the \textbf{mitigation limit} is defined as the value of the generalized bifurcation parameter $\rho$ which separates harmless situations from harmful situation. 	
\end{defn}

As discussed in the literature \cite{Gendelman2010Phys,Gendelman2010SIAM,Bergeot2018}, the mitigation limit (denoted here $\rho_0^*$) is defined, within the zeroth-order approximation, as the value of $\rho$ for which the fixed point $s^e_2$ becomes smaller than $s^\text{U}$ defined by \cref{eq:N2} or for which $s^e_2$ no longer exists (see \cref{eq:exPF}). Indeed, when the largest unstable fixed exists (with, in addition, $s^e_2>s^\text{U}$), it prevents the phase trajectory to grow to infinity on the right branch of the critical manifold.

From \cref{eq:N2,eqPFSFb} the solution of $s^e_2=s^\text{U}$ is derived as
\begin{equation}
\rho^\text{U}_0=a_\mathcal{R}
+\frac{2 \sqrt{1-3 \mu ^2} b_\mathcal{I}+b_\mathcal{I}-\left(2 \sqrt{1-3 \mu ^2}+7\right) \mu ^2 b_\mathcal{I}}{\left(\mu ^2+1\right)^2}
+\frac{\left(3 \mu ^2-4 \sqrt{1-3 \mu ^2}-5\right) \mu  b_\mathcal{R}}{\left(\mu ^2+1\right)^2}.
\label{eq:rho10}
\end{equation}

Then using \eqref{eq:exPF}, the following conditional expression for $\rho_0^*$ is obtained:
\begin{subnumcases}{\label{eq:laidNES}\rho_0^*=}
	\rho^\text{U}_0 \; \text{(\cref{eq:rho10})}, \quad\text{if} \; \mu<\mu^*\label{eq:laidNESa}\\
	\rho^\text{S} \; \text{(\cref{eq:exPF})}, \quad \text{if} \; \mu>\mu^*,\label{eq:laidNESb}
\end{subnumcases}
with $\mu^*$ the special value of $\mu$ solution of $\rho^\text{U}_0=\rho^\text{S}$ which is in general smaller than $1/\sqrt{3}$.

We can notice that within the zeroth-order approximation, the mitigation limit does not depend on the nonlinear parameter $\alpha$ of the NES but only its damping coefficient $\mu$ (together with the parameters of the primary system). Therefore, for given parameters of the primary structure, it exists an optimal value of $\mu$, denoted $\mu_0^\text{opt}$, which maximizes the mitigation limit. Consequently, assuming that $\mu_0^\text{opt}<\mu^*$, $\mu_0^\text{opt}$ is the solution of $\partial_\mu \rho^\text{U}_0=0$. Solving the latter equation yields
\begin{align}
\mu_0^\text{opt}&=-\frac{1}{\mathcal{B}}\Bigg[b_\mathcal{R} b_\mathcal{I}
+\sqrt{\mathcal{B} \left(|b|^4 b_\mathcal{R}^2\right)^{1/3}-3 b_\mathcal{R}^2 \left(b_\mathcal{I}^2+b_\mathcal{R}^2\right)}\nonumber\\
&-\Bigg(-\mathcal{B} \left(|b|^4 b_\mathcal{R}^2\right)^{1/3}-2 \mathcal{B} b_\mathcal{R}^2+2 b_\mathcal{I}^2 b_\mathcal{R}^2
-\frac{2 b_\mathcal{I} b_\mathcal{R} \left(b_\mathcal{I}^2+b_\mathcal{R}^2\right) \left(8 b_\mathcal{I}^2+9 b_\mathcal{R}^2\right)}{\sqrt{\mathcal{B} \left(|b|^4 b_\mathcal{R}^2\right)^{1/3}-3 b_\mathcal{R}^2 \left(b_\mathcal{I}^2+b_\mathcal{R}^2\right)}}\Bigg)^{1/2}\Bigg]
\label{eq:muopt0}
\end{align}
with $\mathcal{B}=3 b_\mathcal{R}^2+4b_\mathcal{I}^2$ and where the subscript $0$ is used to highlight that the expression \eqref{eq:muopt0} is obtained within the zeroth-order approximation.

%-------------------------------------------------------------------------------------------------%
%-------------------------------------------------------------------------------------------------%
% Section
%-------------------------------------------------------------------------------------------------%
%-------------------------------------------------------------------------------------------------%
\section{Scaling law for the slow flow near a fold point}
\label{sec:scallaw}

The Fenichel theory~\cite{Fenichel1979} guarantees the persistence of the relaxation oscillations scenario, described at the end of \cref{sec:respregimes}, for $0<\epsilon\ll 1$ by stating that \cref{eq:slowRealSlow} has an invariant manifold $\mathcal{M}_\epsilon$ in the $O(\epsilon)$-vicinity of $\mathcal{M}_0$ and with the same stability properties with respect to the fast variables as $\mathcal{M}_0$  (attracting or repelling). In this case, concerning the first jump described previously, the jump point at which the trajectory leaves the attracting branch of $\mathcal{M}_\epsilon$ to undergo the jump is no longer the fold point $(s^\text{LF},r^\text{LF})$. The same is true for the arrival point at which the trajectory reaches the right attracting part of $\mathcal{M}_\epsilon$ after the jump which is no longer $(s^\text{U},r^\text{LF})$. The goal of this section is to determine these jump and arrival points and consequently to obtain an analytical prediction of the mitigation limit which, unlike \cref{eq:laidNES}, takes into account $\epsilon$.

To achieve that, one must determine beforehand the scaling law of the slow flow dynamics in the neighborhood of the fold point $(s^\text{LF},r^\text{LF})$, i.e. a law which describes the $\epsilon$-dependance of the distance between $\mathcal{M}_0$ and the actual trajectory of the slow flow for $0<\epsilon\ll 1$. The jump and arrival points are then deduced from this scaling law.

The Fenichel theorem is true for normally hyperbolic branch of $\mathcal{M}_0$ (i.e. outside the fold points) and fails to approach the actual trajectory of the system at fold points.  Indeed, the literature on dynamical systems shows that the trajectory still passes through the region near these fold points but with a nontrivial scaling behavior involving the fractional exponents 1/3 and 2/3 for the perturbation parameter $\epsilon$ (see e.g.~\cite{bk:khuehn2015} Sect. 5.4).

%-------------------------------------------------------------------------------------------------%
% Subsection
%-------------------------------------------------------------------------------------------------%
\subsection{Center manifold reduction of the slow flow at the left fold point}
\label{sec:CMR}

The center manifold reduction technique (as presented for example in~\cite{Guckenheimer_1983Chap3,BerlGen2006}) is used in this section to reduce the slow flow \eqref{eq:slowRealSlow}, in the vicinity of the left fold point $(r^\text{LF},s^\text{LF},\Delta^\text{LF})$, to the normal form of the dynamic saddle-node bifurcation which is then solve to deduce the nontrivial scaling law previously mentioned.

First, assuming $\epsilon=0$ in the right-hand sides of \cref{eq:slowRealSlowa,eq:slowRealSlowb}, one considers the following simplification of the slow flow 
\begin{subequations}
\label{eq:slowRealSlow2}
\begin{align}
	r' &= f(r,s,\Delta)\\
	\epsilon s' &=\tilde{g}_1(r,s,\Delta,0)=g_1(r,s,\Delta),\label{eq:slowRealSlow2a}\\
	\epsilon\Delta' &=\tilde{g}_1(r,s,\Delta,0)=g_2(r,s,\Delta).\label{eq:slowRealSlow2b}
\end{align}
\end{subequations}

The goal in using the center manifold theorem is to reduce the dynamics with respect to the fast variables $s$ and $\Delta$, i.e. obtain an equivalent system which only depends on one variable. To achieve that, the Jacobian matrix ${\bf J_g}$ of (\ref{eq:slowRealSlow2a}-\ref{eq:slowRealSlow2b}) evaluated at the fold point $(r^\text{LF},s^\text{LF},\Delta^\text{LF})$ is computed as
\begin{equation}
{\bf J_g}(r^\text{LF},s^\text{LF},\Delta^\text{LF})=
\begin{pmatrix}
-\frac{\mu }{2}&\frac{1}{9} \left(\sqrt{1-3 \mu ^2}+1\right) \sqrt{\frac{2-\sqrt{1-3 \mu ^2}}{\alpha }}\\
-\frac{3 \left(\sqrt{1-3 \mu ^2}-1\right)}{4 \sqrt{\frac{2-\sqrt{1-3 \mu ^2}}{\alpha }}} &
-\frac{\mu }{2}
\end{pmatrix}.
\label{eq:Jacf}
\end{equation}
which has two real eigenvalues $\lambda_a=0$ and $\lambda_b=-\mu$. 

Then to continue one must obtain polynomial nonlinearities. Indeed, the center manifold reduction requires isolating the linear terms. The functions $g_1$ and $g_2$ are therefore Taylor expanded around the fold point $(r^\text{LF},s^\text{LF},\Delta^\text{LF})$, up to order 2 for the variables $s$ and $\Delta$ and up to order 1 for the variable $r$. Remembering that by definition $g_1(r^\text{LF},s^\text{LF},\Delta^\text{LF})=0$ and $g_2(r^\text{LF},s^\text{LF},\Delta^\text{LF})=0$ that yields
\begin{subequations}
\begin{align}
g_1(r,s,\Delta)&\approx\left(s-s^\text{LF}\right)\partial_{s}g_1(r^\text{LF},s^\text{LF},\Delta^\text{LF})
+\left(\Delta-\Delta^\text{LF}\right)\partial_{\Delta}g_1(r^\text{LF},s^\text{LF},\Delta^\text{LF})
+\tilde{G}_1(r,s,\Delta)\\
g_2(r,s,\Delta)&\approx\left(s-s^\text{LF}\right)\partial_{s}g_2(r^\text{LF},s^\text{LF},\Delta^\text{LF})
+\left(\Delta-\Delta^\text{LF}\right)\partial_{\Delta}g_2(r^\text{LF},s^\text{LF},\Delta^\text{LF})
+\tilde{G}_2(r,s,\Delta).
\end{align}
\label{eq:TaylorG}
\end{subequations}
The polynomial functions $\tilde{G}_1$ and $\tilde{G}_2$ contain nonlinear terms with respect to the $(s-s^\text{LF})$ and $(\Delta-\Delta^\text{LF})$ (only quadratics terms are taken) and linear terms in $(r-r^\text{LF})$.

Denoting ${\bf t}=(s,\Delta)^T$ and ${\bf t}^\text{LF}=(s^\text{LF},\Delta^\text{LF})^T$ and using \cref{eq:TaylorG}, \cref{eq:slowRealSlow2a,eq:slowRealSlow2b} can be written in matrix form as
\begin{equation}
\epsilon  {\bf t}'= \left({\bf t}-{\bf t}^\text{LF}\right) {\bf J_g}(r^\text{LF},{\bf t}^\text{LF})+{\bf \tilde{G}}\left(r-r^\text{LF},{\bf t}-{\bf t}^\text{LF}\right)
\label{eq:slowVar1} 
\end{equation}
where ${\bf \tilde{G}}=(\tilde{G}_1,\tilde{G}_2)^T$. Now, to diagonalize the linear part of \cref{eq:slowVar1} the biorthogonal transformation is used again\footnote{To avoid having to introduce new notations, same symbols as in \cref{sec:red1} are used in this section to describe the biorthogonal transformation except for the subscripts which are now letters ($a$, $b$,...) instead of numbers (1,2,...).}. For that, the right and left eigenvector matrices ${\bf R}=\left[{\bf r}_a \ {\bf r}_b\right]$ and ${\bf L}=\left[{\bf l}_a \ {\bf l}_b\right]$ are considered corresponding to the following right and left eigenvalue problems with respect to ${\bf J_g}(r^\text{LF},{\bf t}^\text{LF})$
\begin{equation}
{\bf J_g}(r^\text{LF},{\bf t}^\text{LF}) {\bf R} = {\bf R} {\bm \Lambda}\quad
\text{and}
\quad
%\begin{equation}
{\bf J_g}(r^\text{LF},{\bf t}^\text{LF})^T {\bf L} = {\bf L} {\bm \Lambda}.
\label{eq:lEVp2}
\end{equation}
where ${\bm \Lambda} = \text{diag}(\lambda_{a}, \lambda_b)= \text{diag}(0,-\mu)$ denotes the diagonal matrix of the eigenvalues. Again the biorthogonal transformation consists of stating
\begin{equation}
{\bf t}-{\bf t}^\text{LF} ={\bf Rq} \quad \text{with} \quad {\bf q=L}^T\left({\bf t}-{\bf t}^\text{LF}\right) 
\label{eq:binormtrans2}
\end{equation}
where ${\bf q}= (q_a,q_b)^T$. Note that here, on a system undergoing a saddle-node bifurcation, the eigenvalues, the eigenvectors and the modal coordinates are real.

The variable $q_a$ is associated to $\lambda_a=0$ and the variable $q_b$ is associated to $\lambda_b=-\mu$. Moreover, the matrices $\bf R$ and $\bf L$ have been chosen such that ${\bf L}^T {\bf R}={\bf I}$ and ${\bf L}^T  {\bf J_g}(r^\text{LF},{\bf t}^\text{LF})\bf R
=\bm \Lambda$. \cref{eq:slowVar1} is then expressed with respect to the variables $q_a$ and $q_b$ as
\begin{equation}
	\epsilon{\bf q}' =  {\bm \Lambda \bf  q} + {\bf G}\left(u,{\bf q}\right)
	\label{eq:MotionEq3bis2}
\end{equation}
where $ {\bf G}\left(u,{\bf q}\right)= {\bf L}^T {\bf \tilde{G}}\left(r-r^\text{LF},{\bf Rq}\right)$ and $u=r-r^\text{LF}$. Because of the latter change of variable the linear terms in $q_a$ and $q_b$ vanish in $ {\bf G}$.

The Jacobian matrix ${\bf J_g}(r^\text{LF},{\bf t}^\text{LF})$ has a null eigenvalue and a negative one. Consequently, the center manifold theorem states that, in the neighborhood of the fold point $(r^\text{LF},s^\text{LF},\Delta^\text{LF})$, studying the following reduced system
\begin{equation}
\epsilon q_a'= G_1\left(q_a,\ell(q_a),u\right)
\end{equation}
is equivalent to studying \cref{eq:MotionEq3bis2}. The function $\ell$ cannot be obtained explicitly and it must therefore be approximated. For sake of simplicity one chooses the \textit{tangent space approximation} (see for example \cite{book:207976}, Chap. 18), i.e. $q_b=\ell(q_a)=0$. 

To obtain a normal form of the dynamic saddle-node bifurcation final approximations are needed. First, because, we study the dynamics near the fold point $(r^\text{LF},s^\text{LF},\Delta^\text{LF})$ we assume that
$f(r,s,\Delta)=f(r^\text{LF},s^\text{LF},\Delta^\text{LF})=f^\text{LF}.$
Then, only terms of order 2 in $q_a$ and of order 1 in $u$ are kept in $G_1$ leading to 
\begin{equation}
G_1 \approx a_1 q_a^2+a_2 u
\end{equation}
with
\begin{equation}
a_1=\frac{3 \left(\sqrt{1-3 \mu ^2}-1\right) \mu ^2+\sqrt{1-3 \mu ^2}+1}{18 \mu ^2}
\quad \text{and} \quad
a_2=\frac{3 \alpha  \sqrt{\frac{2-\sqrt{1-3 \mu ^2}}{\alpha }}}{4 \sqrt{2} \sqrt{3 \mu ^2+\sqrt{1-3 \mu ^2}+1}}.
\label{eq:aa}
\end{equation}

An approximated formulation of the slow flow~\eqref{eq:slowRealSlow2} as the normal form of the dynamic saddle-node bifurcation is then finally obtained as
\begin{subequations}
\label{eq:slowRealSlow3}
\begin{align}
	\hat{\epsilon} q_a'&=h(q_a,v)= q_a^2+v,\label{eq:slowRealSlow3a}\\
	v' &=1 \label{eq:slowRealSlow3b}
\end{align}
\end{subequations}
with
\begin{equation}
v=\frac{a_2}{a_1}u, \quad \hat{\epsilon}=\epsilon f^\text{LF}\frac{a_2}{a_1^2}
\label{eq:vepsh}
\end{equation}
and the following time switching $\tau \rightarrow \frac{a_2 f^\text{LF}}{a_1} \tau$ and where $h(q_a,v)$ satisfies the bifurcation conditions $h(0,0)=0$ and $\partial_{q_a} h(0,0)=0$ and the conditions of the saddle-node bifurcation $\partial_{q_a q_a} h(0,0) \neq 0$ and $\partial_{v} f(0,0) \neq 0$.

The bifurcation is said to be \textit{dynamic} because unlike the \textit{static} saddle-node bifurcation the bifurcation parameter $v$ is a slowly varying parameter (because of \eqref{eq:slowRealSlow3b}).

The critical manifold $\mathcal{M}_{0}$ of \eqref{eq:slowRealSlow3} consists of an attracting branch $\mathcal{M}_{0,\text{a}}=\{(x, y) \in \mathbb{R}^{-}\times \mathbb{R}^{-} \ \; | \; x= -\sqrt{-y}\}$ and a repelling branch $\mathcal{M}_{0,\text{r}}=\{(x, y) \in  \mathbb{R}^{+}\times \mathbb{R}^{-} \ \; | \;  x=\sqrt{-y}\}$. This critical manifold is hyperbolic except in $(0,0)$. Finally, it may be noted that the center manifold reduction amounts to stating that locally the critical manifold is approximated by a parabolic function.

%-------------------------------------------------------------------------------------------------%
% Subsection
%-------------------------------------------------------------------------------------------------%
\subsection{Analytical solution for the normal form of the dynamic saddle-node bifurcation}
\label{sec:ThSol}

The method to solve analytically \cref{eq:slowRealSlow3} is presented for exemple in \cite{BerlGen2006}. It begins by dividing \cref{eq:slowRealSlow3a} by \cref{eq:slowRealSlow3b}, that leads to
\begin{equation}
\hat{\epsilon}\frac{d q_a}{dv}=q_a^2+v.
\end{equation} 

The subsequent scaling $q_a=\hat{\epsilon}^{1 / 3}z$ and $v=-\hat{\epsilon}^{2 / 3} s$ is then introduced to obtain the following equation
\begin{equation}
\frac{d z}{ds}=-z^2+s.
\label{eq:eqZ1}
\end{equation} 
 
Setting $z(s)=\varphi^{\prime}(s) / \varphi(s)$ in \eqref{eq:eqZ1} yields the linear second order equation $\varphi^{\prime \prime}(s)=s \varphi(s)$, whose solution can be expressed in terms of Airy functions as $\varphi(s)=C_1 \mathrm{Ai}(s)+C_2\mathrm{Bi}(s)$ where $C_1$ and $C_2$ are constants. Assuming that $x(-\infty)=-\sqrt{-y}$, only the contribution of $\mathrm{Ai}(s)$ is kept (this is not proved here). Therefore the solution with respect to the original variable $q_a$, denoted $q_a^\star(v)$, is then given by
\begin{equation}
q_a^\star(v)=\hat{\epsilon}^{1/3} \frac{\mathrm{Ai}'\left(-\hat{\epsilon}^{-2/3}v\right)}{\mathrm{Ai}\left(-\hat{\epsilon}^{-2/3}v\right)}.
\label{eq:q1vAi1}
\end{equation}
Note that from \cref{eq:slowRealSlow3} to \cref{eq:q1vAi1} no asymptotic approach has been used.

The first zero of $q_a^\star(v)$ is given by the first zero of the Airy function derivative $\mathrm{Ai}'$ and the first singularity by the first zero of $\mathrm{Ai}$ which are tabulated values (see e.g.~\cite{abramowitz1972handbook}). Therefore, one has
\begin{equation}
q_a^\star(v)=0 \quad \text{for} \quad v=1.01879 \times \hat{\epsilon}^{2/3}
\label{eq:0Aip}
\end{equation}
and
\begin{equation}
q_a^\star(v)\rightarrow\infty \quad \text{for} \quad v=2.33810 \times \hat{\epsilon}^{2/3}.
\label{eq:0Ai}
\end{equation}

\begin{figure}[t!]%\sidecaption
	\centering
	\includegraphics[width=0.55\columnwidth]{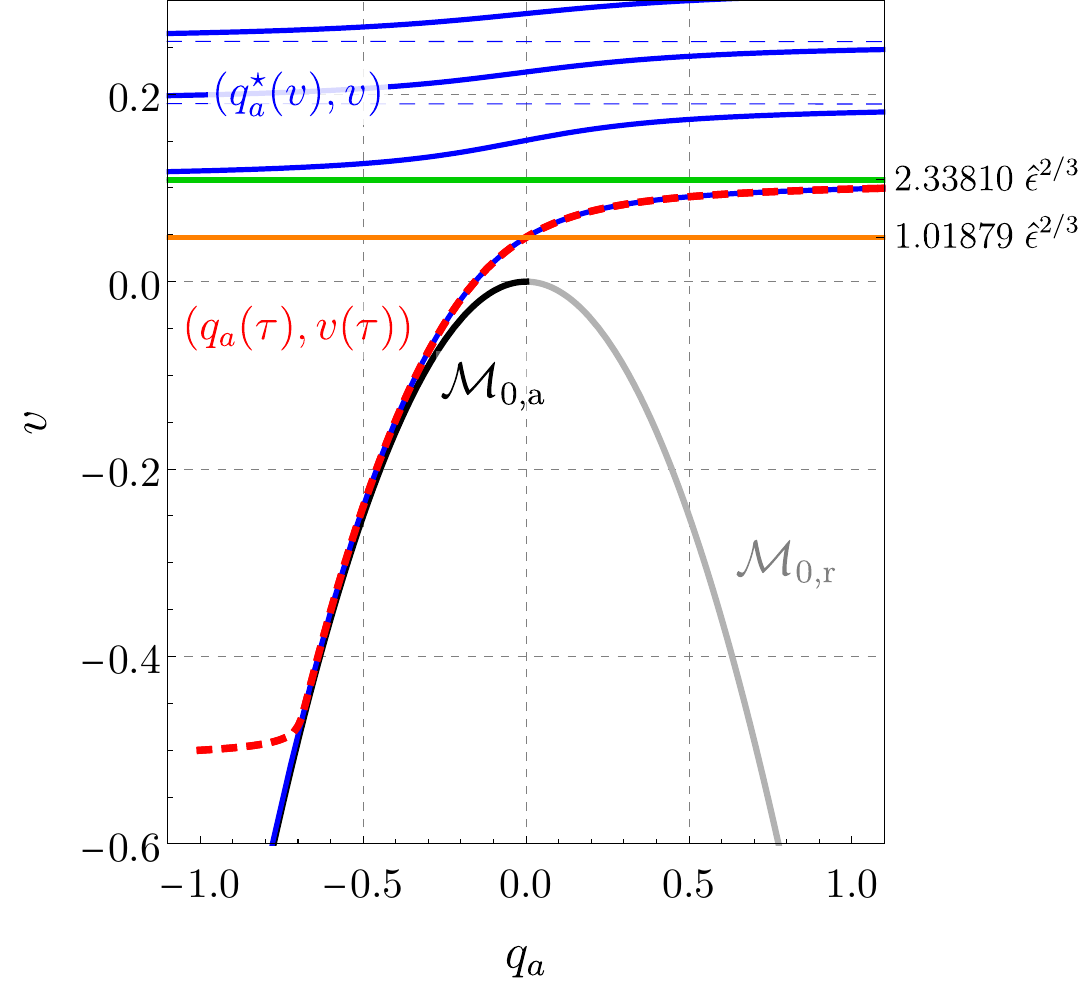}
	\caption{Illustration of the normal form of the dynamic saddle-node bifurcation. Result of the numerical integration of \cref{eq:slowRealSlow3} with initial condition $(q_a(0)=-1,v(0)=-0.5)$ (red dashed line) compared to the analytical scaling law $q_a^\star(y)$ given by \eqref{eq:q1vAi1} (blue line, the dashed parts are the horizontal asymptotes of $q_a^\star(y)$ corresponding to the zeros Airy function) for $\epsilon=0.01$. The first zero and the first singularity of $q_a^\star(v)$ (orange and green lines respectively) and the corresponding critical manifold $\mathcal{M}_{0}$ (attracting part $\mathcal{M}_{0,\text{a}}$ in black and repelling part $\mathcal{M}_{0,\text{r}}$ in gray) are also shown.}
	\label{fig:NormSNBifEx}
\end{figure}

An illustration of the normal form of the dynamic saddle-node bifurcation is shown in \cref{fig:NormSNBifEx}. The direct numerical integration of \cref{eq:slowRealSlow3} with initial condition $(q_a(0)=-1,v(0)-0.5)$ is compared to the analytical solution $q_a^\star(y)$ given by \eqref{eq:q1vAi1} for $\epsilon=0.01$. The first zero, the first singularity of $q_a^\star(v)$ and the corresponding critical manifold $\mathcal{M}_{0}$ (which consists of an attracting part $\mathcal{M}_{0,\text{a}}$ and a repelling part $\mathcal{M}_{0,\text{r}}$) are also shown.

From \cref{eq:binormtrans2,eq:vepsh}, an approximated analytical solution of the slow flow~\eqref{eq:slowRealSlow2} in the neighborhood of the fold point $(r^\text{LF},s^\text{LF},\Delta^\text{LF})$ is obtained as
\begin{equation}
s^\star(r)=s^\text{LF}+{\bf r}_{a,1}\hat{\epsilon}^{1/3} \frac{\mathrm{Ai}'\left(-\hat{\epsilon}^{-2/3}\frac{a_2}{a_1}\left(r-r^\text{LF}\right)\right)}{\mathrm{Ai}\left(-\hat{\epsilon}^{-2/3}\frac{a_2}{a_1}\left(r-r^\text{LF}\right)\right)}
\label{eq:srAi1}
\end{equation}
where $a_1$, $a_2$ and $\hat{\epsilon}$ are given by \eqref{eq:aa} and \eqref{eq:vepsh} respectively and ${\bf r}_{a,1}$ is the first coordinate of the vector~${\bf r}_{a}$.

The previous analysis reveals the nontrivial $\epsilon$-dependence of the slow flow near the fold point, involving the fractional exponents 1/3 and 2/3. Consequently, near the fold point, the trajectory no longer tracks the attracting branch of $\mathcal{M}_{0,\text{a}}$ of the critical manifold at a distance of order $\epsilon$ as it is the case when the trajectory evolves slowly far from the fold point (as predicted by the Fenichel theory in the normally hyperbolic branches of $\mathcal{M}_{0}$).

Two particular values of $r$ can be deduced from \cref{eq:srAi1}: the value at which $s^\star(r)=s^\text{LF}$, denoted $r^0$, and the value for which $s^\star(r)$ tends toward infinity, denoted $r^{\infty}$. These values of $r$ correspond to the first zeros of $\mathrm{Ai}'$ and $\mathrm{Ai}$ respectively (see \cref{eq:0Aip,eq:0Ai}). Therefore, one has
\begin{equation}
s^\star(r)=s^\text{LF}  \quad \text{for} \quad r=r^0=r^\text{LF}+K^0\epsilon^{2/3}
\end{equation}
where $K^0=1.01879 \times \frac{a_1}{a_2}\left(f^\text{LF}\frac{a_2}{a_1^2}\right)^{2/3}$ and
\begin{equation}
s^\star(r)\rightarrow\infty \quad \text{for} \quad
r=r^\infty=r^\text{LF}+K^\infty\epsilon^{2/3}
\label{eq:rinfty}
\end{equation}
where $K^\infty=2.33810 \times \frac{a_1}{a_2}\left(f^\text{LF}\frac{a_2}{a_1^2}\right)^{2/3}$.

%We are particularly interested in $r^\infty$ in next section because it allows us to obtain a new prediction of the mitigation taking into account the value of the perturbation parameter $\epsilon$.

%-------------------------------------------------------------------------------------------------%
% Subsection
%-------------------------------------------------------------------------------------------------%
\subsection{New theoretical prediction of the mitigation limit}
\label{sec:NewAnPred}

%\paragraph{Definiton}

From the results obtained in the previous section, especially \cref{eq:rinfty}, it is possible to obtain a more accurate prediction of the mitigation limit than \eqref{eq:laidNES} which takes into account the value of the small parameter $\epsilon$.

As seen previously (see \cref{sec:predzero}), the mitigation limit is the value of $\rho$ for which the fixed point $s^e_2$ becomes smaller than $s^\text{A}$ (the abscissa of the arrival point of the trajectory on the critical manifold) or for which $s^e_2$ no longer exists (see \cref{eq:exPF}). Within the zeroth-order approximation the arrival point is $s^\text{A}=s^\text{U}$ defined by \eqref{eq:N2}. 

It is not easy to derive a new expression of $s^\text{A}$ from the scaling law~\eqref{eq:srAi1}. However, one can assume that when the trajectory reaches the right attracting par of $\mathcal{M}_0$, the limit value $r^\infty$ has already been reached. Therefore, the solution of $s^e_2=s^\text{A}$, denoted $\rho^\text{U}_\epsilon$, is also the solution of
\begin{subequations}
\label{defNewrhoml}
\begin{align}
H\left(s^e_2(\rho^\text{U}_\epsilon)\right)&=r^{\infty},\\
&=r^\text{LF}+K^{\infty}(\rho^\text{U}_\epsilon)\epsilon^{2/3},\\
&=H(s^\text{LF})+K^{\infty}(\rho^\text{U}_\epsilon)\epsilon^{2/3}.\label{defNewrhomlc}
\end{align}
\end{subequations}

Except for very small values of $\epsilon$ for which $\rho^\text{U}_\epsilon\approx \rho^\text{U}_0$ and the zeroth-order approximation is sufficient, the special value of $\mu$ solution of $\rho_\epsilon^\text{U}=\rho^\text{S}$ is in general larger than $1/\sqrt{3}$, the upper limit of the study interval (see \cref{eq:rocond}). Therefore, we consider here that the mitigation limit is directly $\rho^\text{U}_\epsilon$.

From \cref{defNewrhoml} a new and more accurate value of the optimal value of the NES damping coefficient can be derived. Indeed, denoting $H\left(s^e_2(\rho)\right)=f_1(\rho,\mu)$, $H(s^\text{LF})=f_2(\mu)$ and $K^{\infty}(\rho)\epsilon^{2/3}=f_3(\rho,\mu)$ one has
\begin{equation}
d \rho_\mu=\frac{d_\mu f_2+\partial_\mu f_3-\partial_\mu f_1}
{\partial_\mu f_1-\partial_\mu f_3}.
\label{eq:drho1}
\end{equation}
Therefore solving $d_\mu \rho=0$ together with \cref{defNewrhomlc} gives the coordinates of the maxima of $\rho^\text{U}_\epsilon(\mu)$ and then optimal value of the NES damping coefficient, denoted $\mu^\text{opt}_\epsilon$.

Introducing a new perturbation parameter $\nu=\epsilon^{2/3}$, \cref{defNewrhoml} takes the following form
\begin{equation}
H\left(s^e_2(\rho^\text{U}_\epsilon)\right)=H(s^\text{LF})+\nu K^{\infty}(\rho^\text{U}_\epsilon).
\label{defNewrhomlNu}
\end{equation}
Because $0<\epsilon\ll 1$ one has also $0<\nu\ll 1$, the latter equation is a perturbation of $H\left(s^e_2\right)=H(s^\text{LF})$ which is equivalent to $H\left(s^e_2\right)=H(s^\text{U})$ and consequently to $s^e_2=s^\text{U}$ whose solution is $\rho^\text{U}_0$ given by~\cref{eq:rho10}.

Therefore, to find an explicit formulae for the solution of \eqref{defNewrhoml}, a regular perturbation approach is used by expressing $\rho^\text{U}_\epsilon$ as a power series of $\nu$
\begin{equation}
\rho^\text{U}_\epsilon=\rho^\text{U}_{\epsilon,0}+\nu \rho^\text{U}_{\epsilon,1} +\dots
\label{eq:powerser}
\end{equation}

Then, substituting \eqref{eq:powerser} into \eqref{defNewrhomlNu} and equating the coefficients of identical power of $\nu$ on both sides of the equation lead to the following set of equations (up to the first order)
\begin{subequations}
\label{eq:MethNaive1}
\begin{align}
\boldsymbol{\nu^0:}&&H\left(s^e_2(\rho^\text{U}_{\epsilon,0})\right)&=H(s^\text{LF})\label{eq:MethNaive1a}\\
\boldsymbol{\nu^1:}&& \rho^*_1 d_s H\left(s^e_2(\rho^\text{U}_{\epsilon,0})\right)d_\rho s^e_2(\rho^\text{U}_{\epsilon,0})&=K^{\infty}(\rho^\text{U}_{\epsilon,0})\label{eq:MethNaive1b}
\end{align}
\end{subequations} 
As expected, from \eqref{eq:MethNaive1a} one has $\rho^\text{U}_{\epsilon,0}=\rho^\text{U}_0$ (see \cref{eq:rho10}) and therefore \cref{eq:MethNaive1b} yields
\begin{equation}
\rho^\text{U}_{\epsilon,1}=\frac{K^{\infty}(\rho^\text{U}_0)}{d_s H\left(s^e_2(\rho^\text{U}_0)\right)d_\rho s^e_2(\rho^\text{U}_0)}.
\end{equation}

Finally, to a first approximation, one has the following asymptotic expression for $\rho^\text{U}_\epsilon$
\begin{equation}
\rho^\text{U}_\epsilon=\rho^\text{U}_0+\epsilon^{2/3}\frac{K^{\infty}(\rho^\text{U}_0)}{d_s H\left(s^e_2(\rho^\text{U}_0)\right)d_\rho s^e_2(\rho^\text{U}_0)}.
\label{eq:mitlimEps}
\end{equation}

The asymptotic series \eqref{eq:powerser} converges if the functions of $\rho$ in \cref{defNewrhomlNu} (i.e. $s_2^e(\rho)$ and $K^{\infty}(\rho)$) can be expanded as a convergent Taylor series at $\rho=\rho^\text{U}_0$. Unfortunately $s_2^e(\rho)$ has an infinite derivative at $\rho=\rho^\text{S}$ (see \cref{eq:exPF}). Because of that, at $\mu=\mu^*$ where $\rho^\text{U}_0=\rho^\text{S}$ the term $\rho^\text{U}_{\epsilon,1}$ in the series~\cref{eq:powerser} is equal to zero but the following terms grow to infinity making the series divergent. However, except at $\mu=\mu^*$ and for sufficiently small values of $\epsilon$, the expression \eqref{eq:mitlimEps} shows again a nontrivial $\epsilon$-dependence stating that $\rho^\text{U}_\epsilon$ is a $O(\epsilon^{2/3})$-perturbation of $\rho^\text{U}_0$.

Equating to zero the derivative, with respect to $\mu$, of the right-hand side of \cref{eq:mitlimEps} and performing again a perturbative technique with respect to $\nu=\epsilon^{2/3}$, an approximative expression of $\mu^\text{opt}_\epsilon$ is obtained as
\begin{equation}
\mu^\text{opt}_\epsilon=\mu^\text{opt}_0+\epsilon^{2/3}\frac{d_\mu\left(\frac{K^{\infty}(\rho^\text{U}_0)}{d_s H\left(s^e_2(\rho^\text{U}_0)\right)d_\rho s^e_2(\rho^\text{U}_0)}\right)_{\mu=\mu^\text{opt}_0}}{d_{\mu\mu}\left(\rho^\text{U}_0\right)_{\mu=\mu^\text{opt}_0}}
\label{eq:muoptEpsAs}
\end{equation}
where $\mu^\text{opt}_0$ is given by \cref{eq:muopt0}. Obviously, the nontrivial $\epsilon$-dependence is found again as, if $\epsilon$ is sufficiently small, $\mu^\text{opt}_\epsilon$ is a $O(\epsilon^{2/3})$-perturbation of $\mu^\text{opt}_0$. Of course the limitations previously mentioned for \cref{eq:mitlimEps} are also valuable for \cref{eq:muoptEpsAs}.

%-------------------------------------------------------------------------------------------------%
%-------------------------------------------------------------------------------------------------%
% Section
%-------------------------------------------------------------------------------------------------%
%-------------------------------------------------------------------------------------------------%
\section{Application to an aeroelastic aircraft wing system coupled to an NES}
\label{sec:appl}

The methodology presented in previous sections is illustrated here on an aeroelastic aircraft wing system coupled to an NES (see \cref{fig:Airfoil}) which has already been used in the past in several works about LCO mitigation by means of one or several NESs~\cite{LeeAIAApI2007,LeeAIAApI2007b,Gendelman2010SIAM,bergeot2019}. First, equations of motion the wing-NES system are presented in \cref{sec:AirfoilNESmod}. Then, the comparison between theoretical results and numerical simulations is performed in \cref{sec:compsec}.

%-------------------------------------------------------------------------------------------------%
% Subsection
%-------------------------------------------------------------------------------------------------%
\subsection{Equations of the aircraft wing model coupled to one NES}
\label{sec:AirfoilNESmod}

\begin{figure}[t!]
	\centering
	\includegraphics[width=0.85\columnwidth]{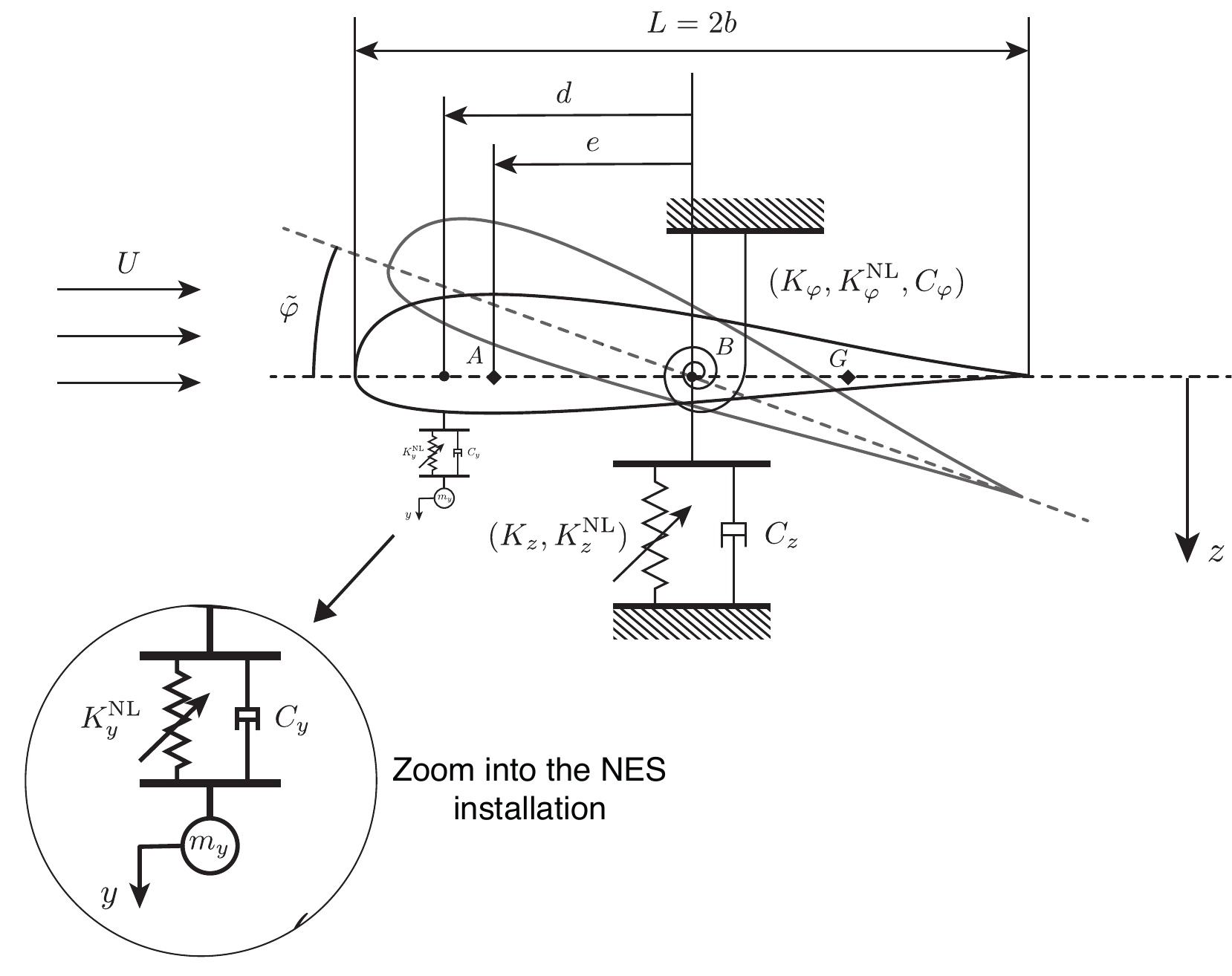}
	\caption{Sketch of the two DOFs aircraft wing coupled to one NES. $z$ and $\tilde{\varphi}$ are respectively the heave and the angle of attack (pitch) of the wing and $y$ is the displacement of the NES. $A$ is the aerodynamic center, $B$ the elastic axis, $G$ the center of gravity of the aircraft wing. $e$ is the location aerodynamic center $A$ measured from $B$ (positive ahead of $B$). $K_z$ and $K_\varphi$ are the linear heave and pitch stiffnesses respectively whereas $K_z^\text{NL}$ and $K_\varphi^\text{NL}$ are the cubic heave and pitch stiffnesses. $C_z$ and $C_\varphi$ are the heave and pitch damping coefficients. $U$ is the constant and uniform flow speed around the wing and $d$ is the offset attachment of the NES to the wing, also measured from $B$.}
	\label{fig:Airfoil}
\end{figure}

The equations of motion of the model presented in \cref{fig:Airfoil} can be derived for example by means of the Lagrange equations. Assuming a small angle between the wing and the horizontal, the Lagrange equations yield the following equations of motion with respect to the heave $z$ and the angle of attack (pitch) $\tilde{\varphi}$ of the wing and to the NES displacement $y$ (see \cref{fig:Airfoil})
\begin{subequations}
	\label{eq:airfioldim}
	\begin{align}
	m \ddot{z}+S_{\varphi}\ddot{\tilde{\varphi}}+K_z z+C_z \dot{z}+
	q P_a\Gamma\left(\tilde{\varphi}+\dot{z}/U \right)+K_z^\text{NL} z^3&+\nonumber\\
	C_{y}\big(\dot{z}-d \dot{\tilde{\varphi}} -\dot{y}\big) +K_y^\text{NL}\left(z-d \tilde{\varphi} -y\right)^3&=0\\
	I_{\varphi}\ddot{\tilde{\varphi}}+S_{\varphi} \ddot{z}+K_\varphi \tilde{\varphi}+C_\varphi \dot{\tilde{\varphi}}-
	q e P_a\Gamma\left(\tilde{\varphi}+\dot{z}/U \right)+K_\varphi^\text{NL} \tilde{\varphi}^3&+\nonumber\\
	d C_{y}\big(\dot{z}-d \dot{\tilde{\varphi}} -\dot{y}\big) +d K_y^\text{NL}\left(z-d \tilde{\varphi} -y\right)^3&=0\\
	m_{y}\ddot{y} +C_{y}\big(d \dot{\tilde{\varphi}} +\dot{y}-\dot{z}\big) +K_y^\text{NL}\left(d \tilde{\varphi} +y-z\right)^3&=0
	\end{align}
\end{subequations}
where again the time derivative is denoted $``\;^.\;"$. $b=L/2$ is the semichord length. $A$ is the aerodynamic center, $B$ the elastic axis, $G$ the center of gravity of the aircraft wing. $e$ is the location aerodynamic center $A$ measured from $B$ (positive ahead of $B$). The quantities $m$ and $I_\varphi$ are respectively the mass of the wing and its moment of inertia with respect to $B$. $S_{\varphi}=m BG$ is the mass unbalance in the wing. $K_z$ and $K_\varphi$ are the linear heave and pitch stiffnesses respectively whereas $K_z^\text{NL}$ and $K_\varphi^\text{NL}$ are the cubic heave and pitch stiffnesses. $C_z$ and $C_\varphi$ are the heave and pitch damping coefficients. $U$ is the constant and uniform flow speed around the wing and $q=\frac{1}{2}\rho_{\infty}U^2$ is the dynamic pressure where $\rho_{\infty}$ is the density of the flow. $P_a$ is the planform of the wing, $\Gamma$ is the lift curve slope and $d$ is the offset attachment of the NES to the wing. Finally, $m_{y}$, $C_{y}$ and $K_y^\text{NL}$ ($m=1,\dots,M$) are the mass, the damping coefficient and the cubic stiffness of the NES respectively. 

For convenience, \cref{eq:airfioldim} is written in a dimensionless form as follows
\begin{subequations}
	\label{eq:airfiolnodim}
	\begin{align}
	\tilde{x}''+s_{\varphi}\tilde{\varphi}''+\Omega^2 \tilde{x}+\zeta_x \tilde{x}'+
	\gamma\Theta\left(\Theta\tilde{\varphi}+\tilde{x}' \right)+
	\tilde{\xi}_x \tilde{x}^3+\tilde{m}_{h}\tilde{h}''&=0\\
	r_{\varphi}^2\tilde{\varphi}''+s_{\varphi} \tilde{x}''+r_\varphi^2 \tilde{\varphi}+\zeta_\varphi \tilde{\varphi}'-
	\eta \gamma  \Theta\left(\Theta\varphi+\tilde{x}' \right)+\tilde{\xi}_\varphi \tilde{\varphi}^3+\delta\tilde{m}_{h}\tilde{h}''&=0\\
	\tilde{m}_{h}\tilde{h}'' + \tilde{\zeta}_{h}\big(\delta \tilde{\varphi}' +\tilde{h}'-\tilde{x}'\big) +\tilde{\xi}_{h}(\delta \tilde{\varphi} +\tilde{h}-\tilde{x})^3&=0 \label{eq:airfiolnodimc}
	\end{align}
\end{subequations}
where the time $t$ has been replaced by the dimensionless time $t'=\omega_{\varphi} t$ (with $\omega_{\varphi}=\sqrt{K_{\varphi}/I_{\varphi}}$) and time derivative $d_{t'}$ is denoted $"\;'\;"$. The dimensionless displacements are defined by $\tilde{x}=z/b$, $\tilde{h}=y/b$. Moreover, $\delta=d/b$, $s_{\varphi}=S_{\varphi}/(mb)$, $\Omega=\omega_z/\omega_{\varphi}$ (with $\omega_z=\sqrt{K_z/m}$) and $\eta=e/b$. $r_\varphi=\sqrt{I_{\varphi}/(mb^2)}$ is the radius of gyration of the cross section of the wing. The dimensionless nonlinear stiffnesses and damping coefficients are $\tilde{\xi}_x=K_z^\text{NL} b^2/(m\omega^2_\varphi)$, $\zeta_x=C_z/(m \omega_\varphi)$, $\tilde{\xi}_\varphi=K_\varphi^\text{NL}/(mb^2 \omega^2_\varphi)$, $\zeta_\varphi=C_\varphi/(m b^2 \omega_\varphi)$, $\tilde{\xi}_{h}=K_y^\text{NL} b^2/(m\omega^2_\varphi)$, $\tilde{\zeta}_{h}=C_{y}/(m \omega_\varphi)$. The mass and density ratios are defined by $\tilde{m}_{h}=m_{y}/(m\omega_\varphi^2)$ and $\gamma=b P_a \rho_\infty\Gamma/(2m)$ respectively. Finally, the bifurcation parameter under consideration is the reduced speed of the flow $\Theta=U/(b\omega_\varphi)$.

\Cref{eq:airfiolnodim} has the general form given by \cref{eq:systprotec0} with $\tilde{\bf x}=(\tilde{x},\tilde{\varphi})^T$ and
\begin{equation}
	\label{eq:matex}
\begin{tabular}{ccc}
	${\bf \tilde{M}}=
	\begin{pmatrix}
	1 &  s_\varphi \\
	s_\varphi  & r_\varphi^2
	\end{pmatrix}$, &
	$\bf \tilde{C} =
	\begin{pmatrix}
	\zeta_x+\gamma \Theta & 0\\
	-\eta \gamma \Theta &  \zeta_{\varphi}
	\end{pmatrix}$, &
	${\bf  \tilde{K}} =
	\begin{pmatrix}
	\Omega^2 &  \gamma \Theta^2\\
	0 & r_\varphi^2-\eta \gamma \Theta^2
	\end{pmatrix}$,\\
	${\bf \tilde{A}}=(1,-\delta)$, &
	${\bf \tilde{B}}=(1,-\delta)^T$, &
	${\bf \tilde{G}^{NL}}=(\tilde{\xi}_x\tilde{x}^3,\tilde{\xi}_\varphi\tilde{\varphi}^3)^T$.
	\end{tabular}
\end{equation}

The expression of the nonlinear vector function is ${\bf \tilde{g}^{NL}}=\left(\tilde{\xi}_x\tilde{x}^3,\tilde{\xi}_\varphi\tilde{\varphi}^3\right)^T$. Moreover, one defines rescaled parameters and variables as
$\tilde{\xi}_{x} = \epsilon \xi_{x}$, $\tilde{\xi}_{\varphi} = \epsilon  \xi_{\varphi}$, 
$\tilde{m}_{h} = \epsilon m_{h}$, $\tilde{\zeta}_{h} = \epsilon  \zeta_{h}$, 
$x=\frac{\tilde{x}}{\sqrt{\epsilon}}$, $\varphi=\frac{\tilde{\varphi}}{\sqrt{\epsilon}}$ and $h=\frac{\tilde{h}}{\sqrt{\epsilon}}$
to obtain the final form of the equations of motion
\begin{subequations}
	\label{eq:airfiolnodim2a}
	\begin{align}
	x''+s_{\varphi}\varphi''+\Omega^2 x+\zeta_x x'+
	\gamma\Theta\left(\Theta\varphi+x' \right)+ \epsilon m_{h}h''&=0\\
	r_{\varphi}^2\varphi''+s_{\varphi} x''+r_\varphi^2 \varphi+\zeta_\varphi \varphi'-
	\eta \gamma \Theta\left(\Theta\varphi+x' \right)+\epsilon\delta m_{h}h''&=0\\
	h'' + \zeta_{h} \big(\delta \varphi' +h'-x'\big) +\xi_h\left(\delta \varphi +h-x\right)^3&=0
	\label{eq:airfiolnodimc2a}
	\end{align}
\end{subequations}
with $\zeta_{h}=\tilde{\zeta}_{h}/m_{h}$ and $\xi_h=\tilde{\xi}_{h}/m_{h}$.

The physical bifurcation parameter under consideration is the reduced speed of the air flow $\Theta$ and the following parameters for the primary system are used
\begin{equation}
\label{eq:airfoilpar}
\begin{split}
\zeta_x&=0.01, \quad \zeta_\varphi=0.01, \quad s_\varphi=0.2, \quad r_\varphi=0.5,\\
\Omega&=0.5, \quad \eta=0.4, \quad \text{and} \quad \gamma=0.2. 
\end{split}
\end{equation}

With the parameters \eqref{eq:airfoilpar}, the aircraft wing model undergoes an Hopf bifurcation at $\Theta^\text{wo}_\text{Hopf}=0.933$ at which the real part of the first eigenvalue of \cref{eq:airfiolnodim2a} vanishes, i.e. $\text{Re}(\lambda_1)=0$, and then becomes positive. It is recalled here that the pulsation $\omega$ and the generalized bifurcation parameter $\rho$ are defined from the eigenvalue $\lambda_1$ of the unstable mode by \cref{eq:rhodef}.

The NES setup is defined by
\begin{equation}
m_h=1, \quad \delta=-0.8, \quad \xi_h=5
\label{eq:NESpar}
\end{equation}
and $\zeta_h$ and $\epsilon$ which are specified below depending of the considered example.

\begin{figure*}[t!]
	\centering
	\subfigure[]{\includegraphics[width=0.47\columnwidth]{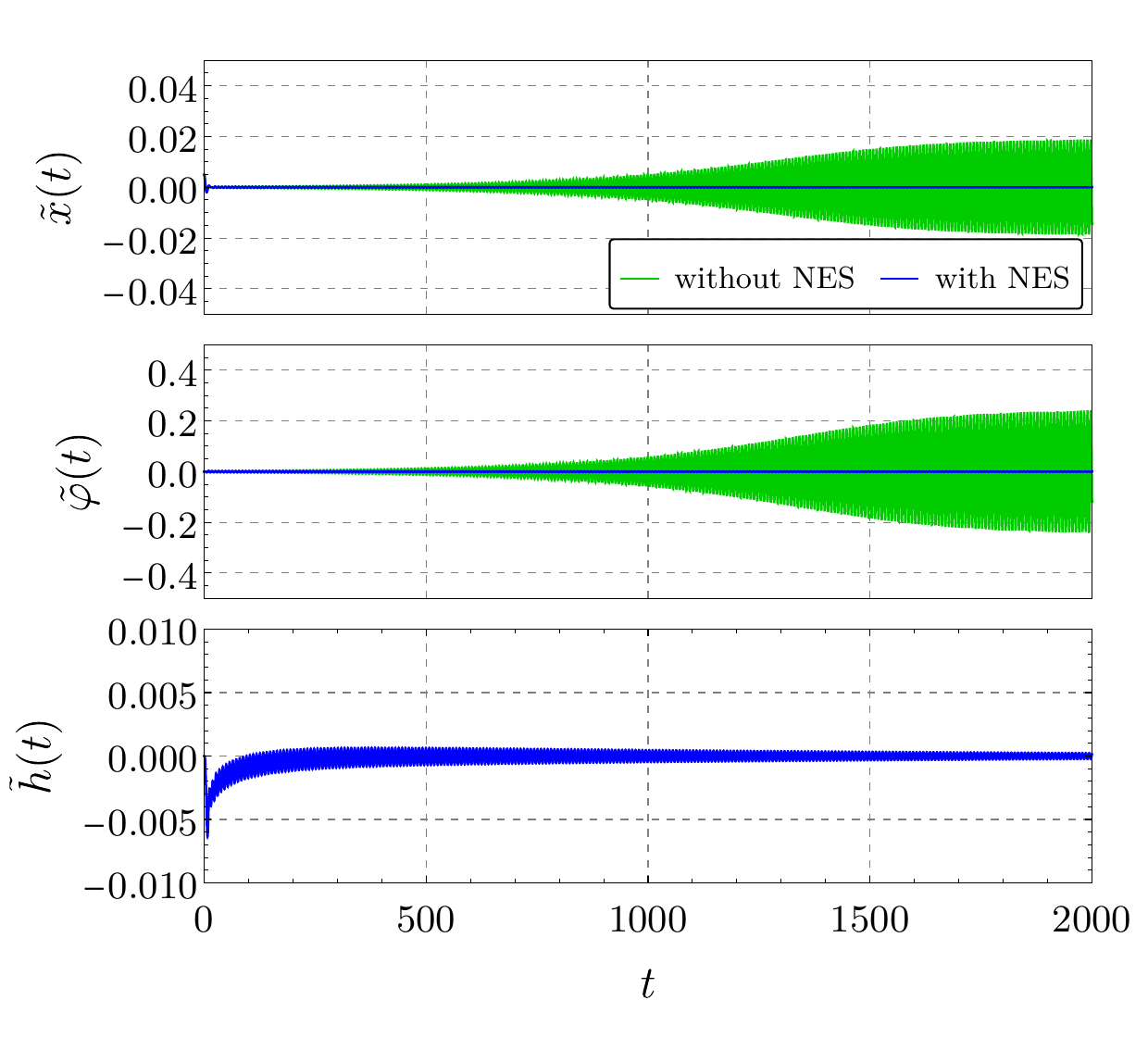}\label{fig:pFlutterExSimua}}
	\subfigure[]{\includegraphics[width=0.47\columnwidth]{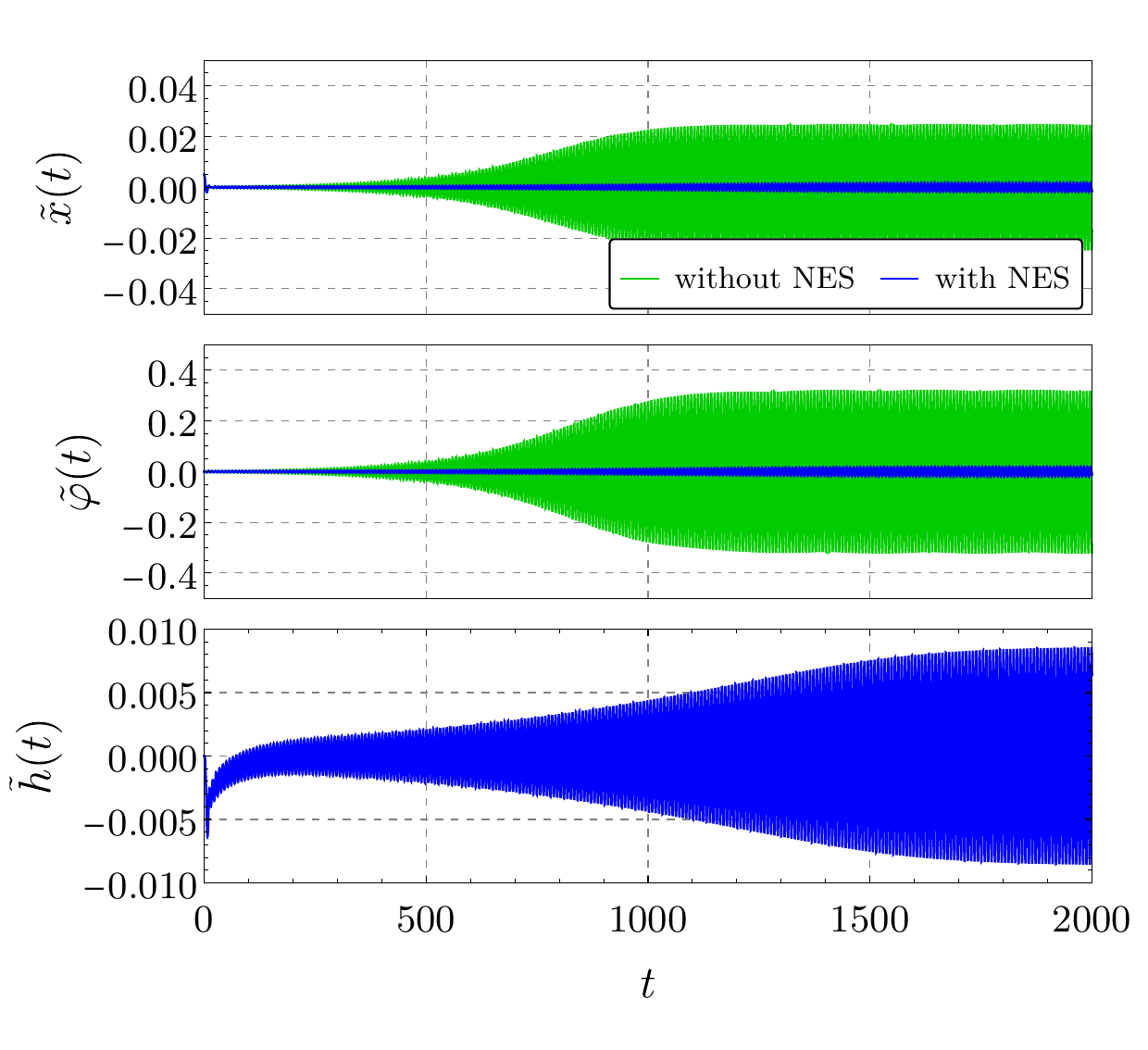}\label{fig:pFlutterExSimub}}
	\subfigure[]{\includegraphics[width=0.47\columnwidth]{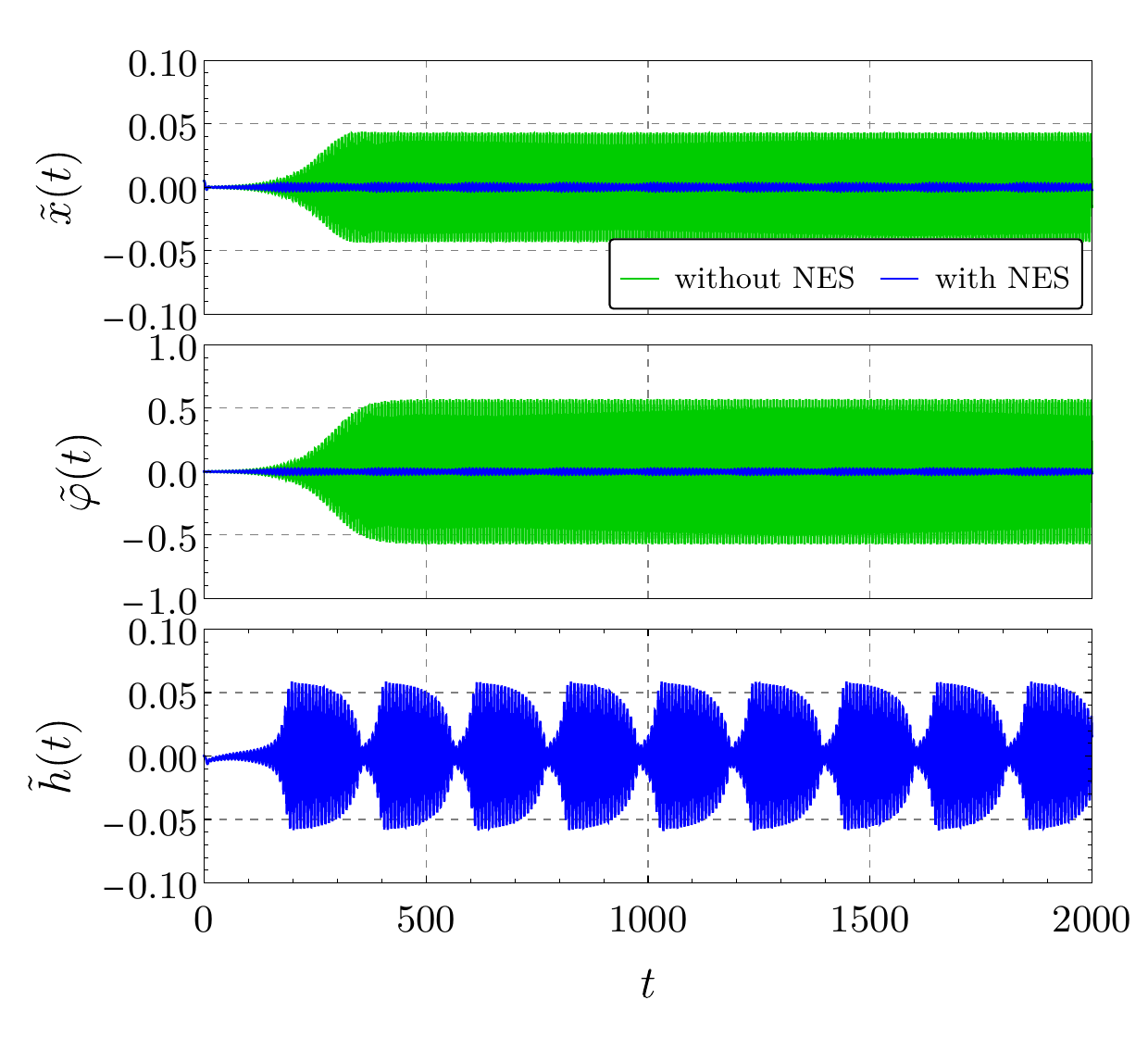}\label{fig:pFlutterExSimuc}}
	\subfigure[]{\includegraphics[width=0.47\columnwidth]{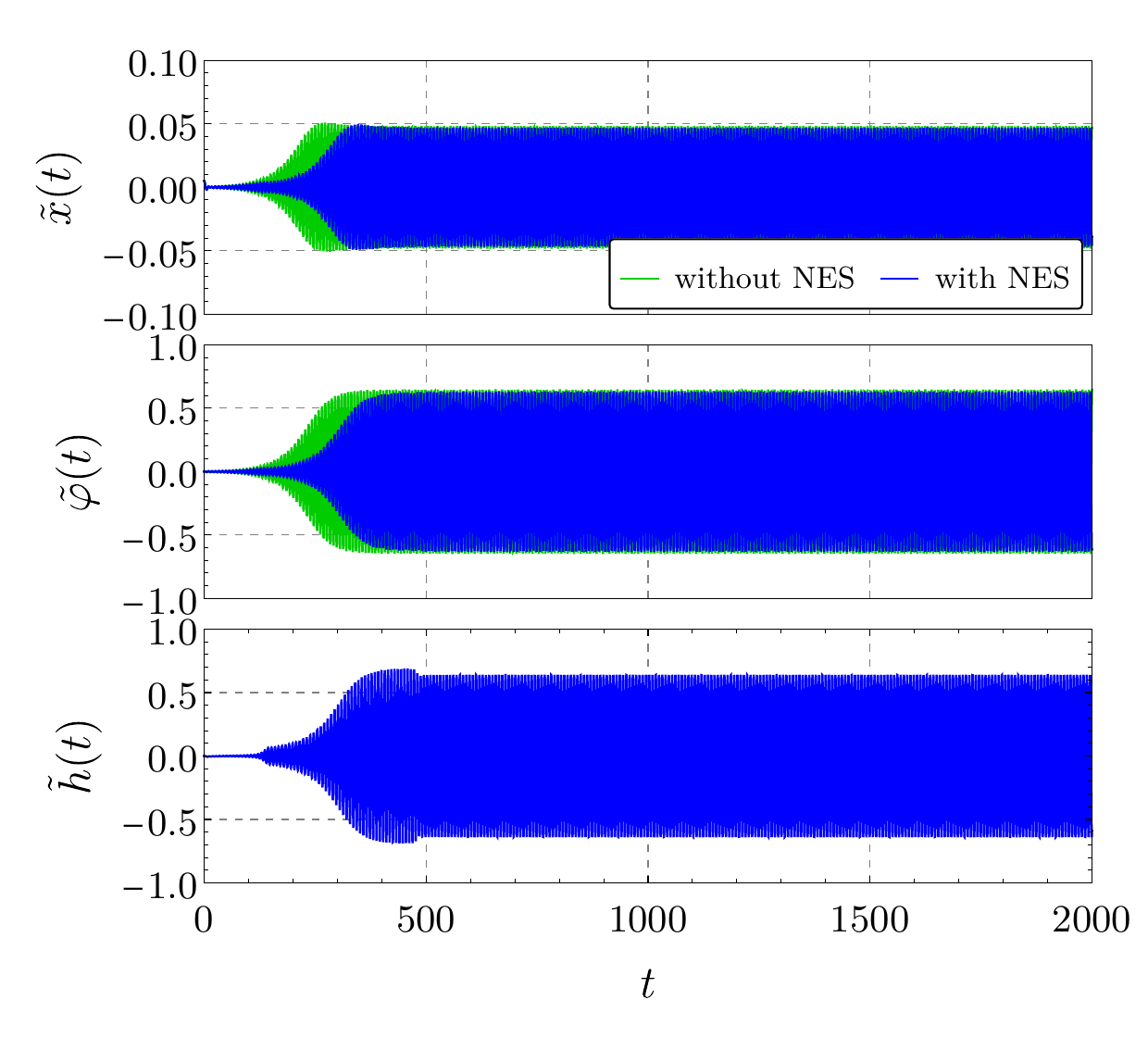}\label{fig:pFlutterExSimud}}
	\caption{Direct numerical integration of the wing-NES system \eqref{eq:airfiolnodim} (in blue) depicting the several response regimes described in \cref{sec:respregimes} with, in addition to \eqref{eq:airfoilpar} and \eqref{eq:NESpar}, $\epsilon=0.01$, $\zeta_h=0.3$, $\xi_h=4$, $\xi_x=4$ and $\xi_\varphi=8$. Four values of  the reduced speed are used, namely: (a) $\Theta=0.94$, (b) $\Theta=0.945$, (c) $\Theta=0.97$ and (d) $\Theta=0.98$ corresponding respectively to complete suppression, mitigation through a periodic regime, mitigation through a SMR and finally no mitigation of the harmful limit cycle oscillations of the primary structure. For comparison purposes, for each situation, a direct numerical integration of the wing system without NES is also shown (in green).}
	\label{fig:pFlutterExSimu}
\end{figure*}

The several response regimes described in \cref{sec:respregimes} are illustrated here by means of numerical simulations of the wing-NES system \eqref{eq:airfiolnodim} (see \cref{fig:pFlutterExSimu}) with, in addition to \eqref{eq:airfoilpar} and \eqref{eq:NESpar}, $\epsilon=0.01$, $\zeta_h=0.3$, $\xi_x=4$ and $\xi_\varphi=8$. Four values of  the reduced speed are used, namely: $\Theta=0.94$ (see \cref{fig:pFlutterExSimua}), $\Theta=0.945$ (see \cref{fig:pFlutterExSimub}),  $\Theta=0.97$ (see \cref{fig:pFlutterExSimuc}) and $\Theta=0.98$ (see \cref{fig:pFlutterExSimud}) corresponding respectively to complete suppression, mitigation through a periodic regime, mitigation through a SMR and finally no mitigation of the harmful LCOs of the primary structure. For comparison purposes, for each situation, a direct numerical integration of the aircraft wing system without NES is also shown.

%-------------------------------------------------------------------------------------------------%
% Subsection
%-------------------------------------------------------------------------------------------------%
\subsection{Comparison between theoretical results and numerical simulations}
\label{sec:compsec}

The slow flow dynamics of \cref{eq:airfiolnodim2a}, its critical manifold $\mathcal{M}_0$ \eqref{eq:CM00}, the zeroth-order analytical expressions of the mitigation limit \eqref{eq:laidNES} and of the optimal value of NES damping coefficient~\eqref{eq:muopt0} are derived following the method described in \cref{sec:slowflow,sec:critman,sec:predzero} respectively. Note that through \cref{eq:parammualp}, the parameters $\mu$ and $\alpha$ are here defined as $\mu=\zeta_h/\omega$ and $\alpha=\xi_h/\omega^2$. Simultaneously, the scaling law of the trajectory in the neighborhood of the left fold point of $\mathcal{M}_0$, i.e. \cref{eq:srAi1}, is computed. Afterwards, solving \cref{defNewrhoml} and finding the root of \cref{eq:drho1}, $\rho^\text{U}_\epsilon(\mu)$ and $\mu^\text{opt}_\epsilon$ are obtained. Finally, the asymptotic analytical expressions of the latter, given by \cref{eq:mitlimEps,eq:muoptEpsAs} respectively, are computed.

\begin{figure}[t!]%\sidecaption
	\centering
	\subfigure[]{\includegraphics[width=0.57\columnwidth]{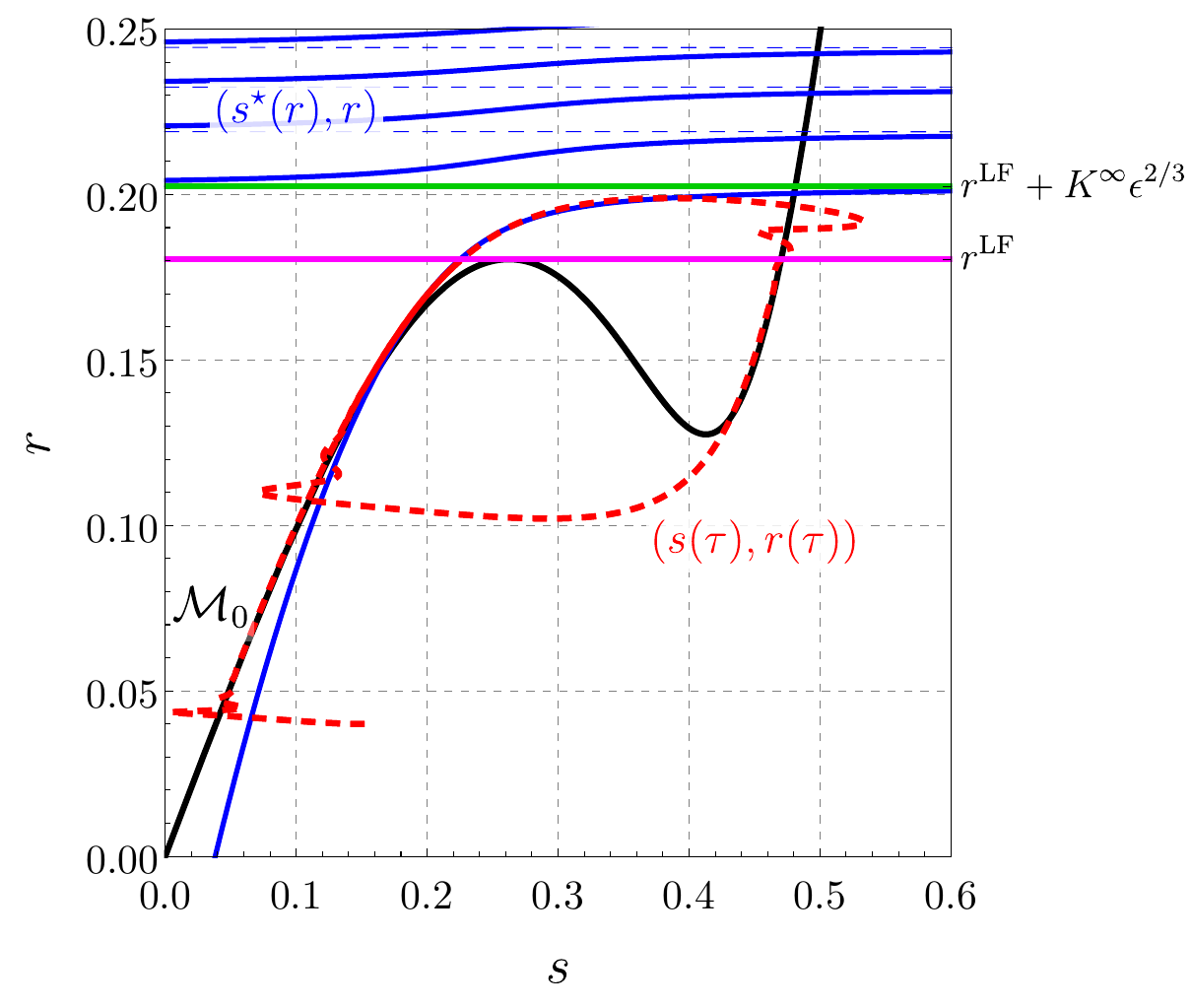}\label{fig:pExAirfoila}}
	\subfigure[]{\raisebox{1cm}{\includegraphics[width=0.42\columnwidth]{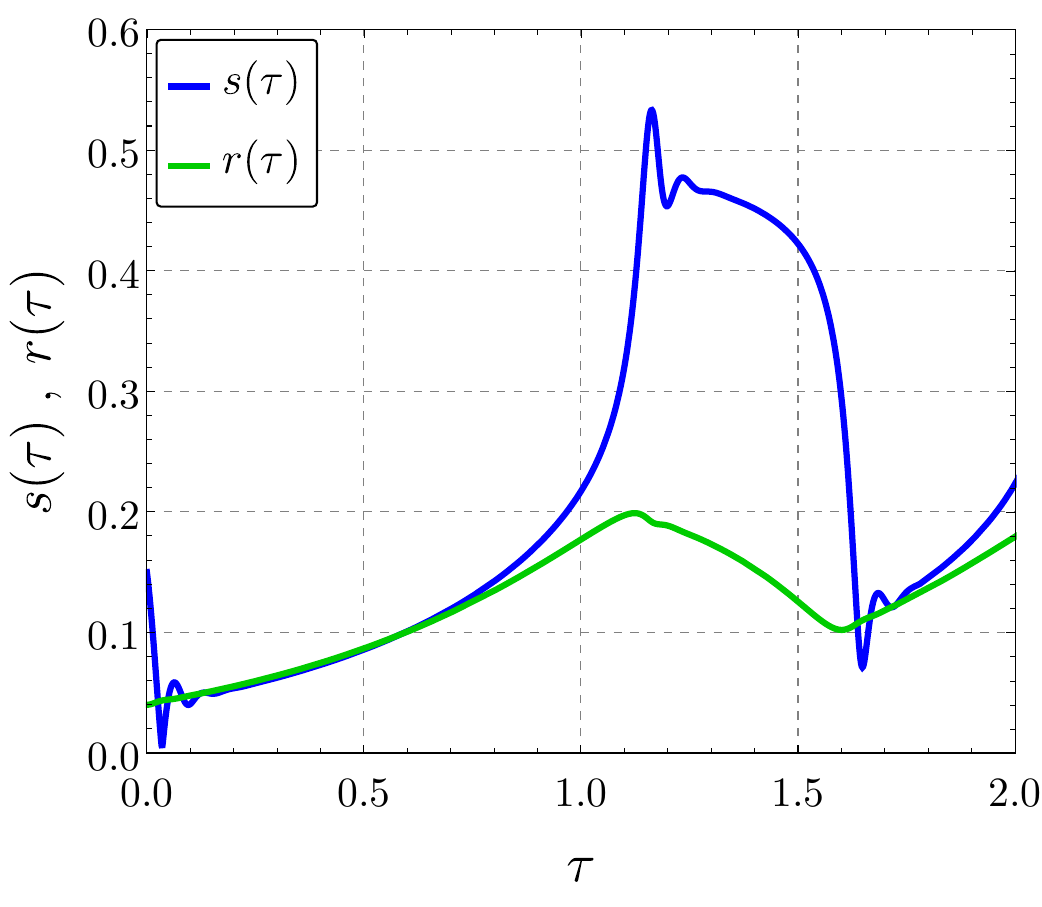}\label{fig:pExAirfoilb}}}
	\caption{Illustration of the proposed analytical procedure. (a) The critical manifold $\mathcal{M}_0$ \eqref{eq:CM00} (black line) superimposed to the direct numerical integration of the slow flow \eqref{eq:slowRealSlow2} (red line) and its scaling law near the left fold point given by \cref{eq:srAi1} (blue line, the dashed parts are the horizontal asymptotes of $s^\star(r)$ due to the zeros of the Airy function). The values of $r^\text{LF}=H(s^\text{LF})$ (see \cref{eq:rNrm}) and $r^\infty=r^\text{LF}+K^\infty\epsilon^{2/3}$ (see \cref{eq:rinfty}) are depicted by magenta and green horizontal lines respectively. (b) Times series of $s(\tau)$ (blue line) and $r(\tau)$ (green line) obtained from the numerical integration of the slow flow \eqref{eq:slowRealSlow2}. The times series are focused on the first relaxation cycle. Parameters \eqref{eq:airfoilpar} and \eqref{eq:NESpar} are used with $\zeta_h=0.25$, $\epsilon=0.005$ and $\Theta=0.95$.}
	\label{fig:pExAirfoil}
\end{figure}

To illustrate the proposed analytical procedure, \cref{fig:pExAirfoila} shows the critical manifold $\mathcal{M}_0$ \eqref{eq:CM00}, superimposed to the direct numerical integration of the slow flow \eqref{eq:slowRealSlow2} (see also \cref{fig:pExAirfoilb} for the times series of $s(\tau)$ and $r(\tau)$ focused on the first relaxation cycle) and its scaling law given by \cref{eq:srAi1}. The values of $r^\text{LF}=H(s^\text{LF})$ (see \cref{eq:rNrm}) and $r^\infty=r^\text{LF}+K^\infty\epsilon^{2/3}$ (see \cref{eq:rinfty}) are depicted by horizontal lines. The numerical integration of the slow flow (see red dashed line in \cref{fig:pExAirfoila}) shows a relaxation oscillations scenario in $(s,r)$-plane. Indeed, the trajectory of undergoes a succession of almost horizontal fast parts and slow parts evolving in the $O(\epsilon)$-vicinity of the critical manifold $\mathcal{M}_0$ (here, at the scale of the figure and apart from the fold points, the trajectory and $\mathcal{M}_0$ seem superimposed). The first horizontal fast part is from the initial condition $(s(0)=0.15,r(0)=0.04)$ to the left attracting part of $\mathcal{M}_0$ and the other are fast jumps between the left and right attracting parts of $\mathcal{M}_0$ as described in \cref{sec:respregimes} and in the introduction of \cref{sec:scallaw}.

The figure reveals a good agreement between the numerical simulation of the slow flow and its analytical scaling law in the neighborhood of the left fold point. One can see also that choosing $r^\infty$ as the ordinate of the point of intersection between the trajectory of the slow flow and the right attracting branch of $\mathcal{M}_0$  is a more accurate approximation than choosing $r^\text{LF}$ as this is the case within the zeroth-order approximation. 

\begin{figure*}[t!]
	\centering
	\includegraphics[width=0.9\columnwidth]{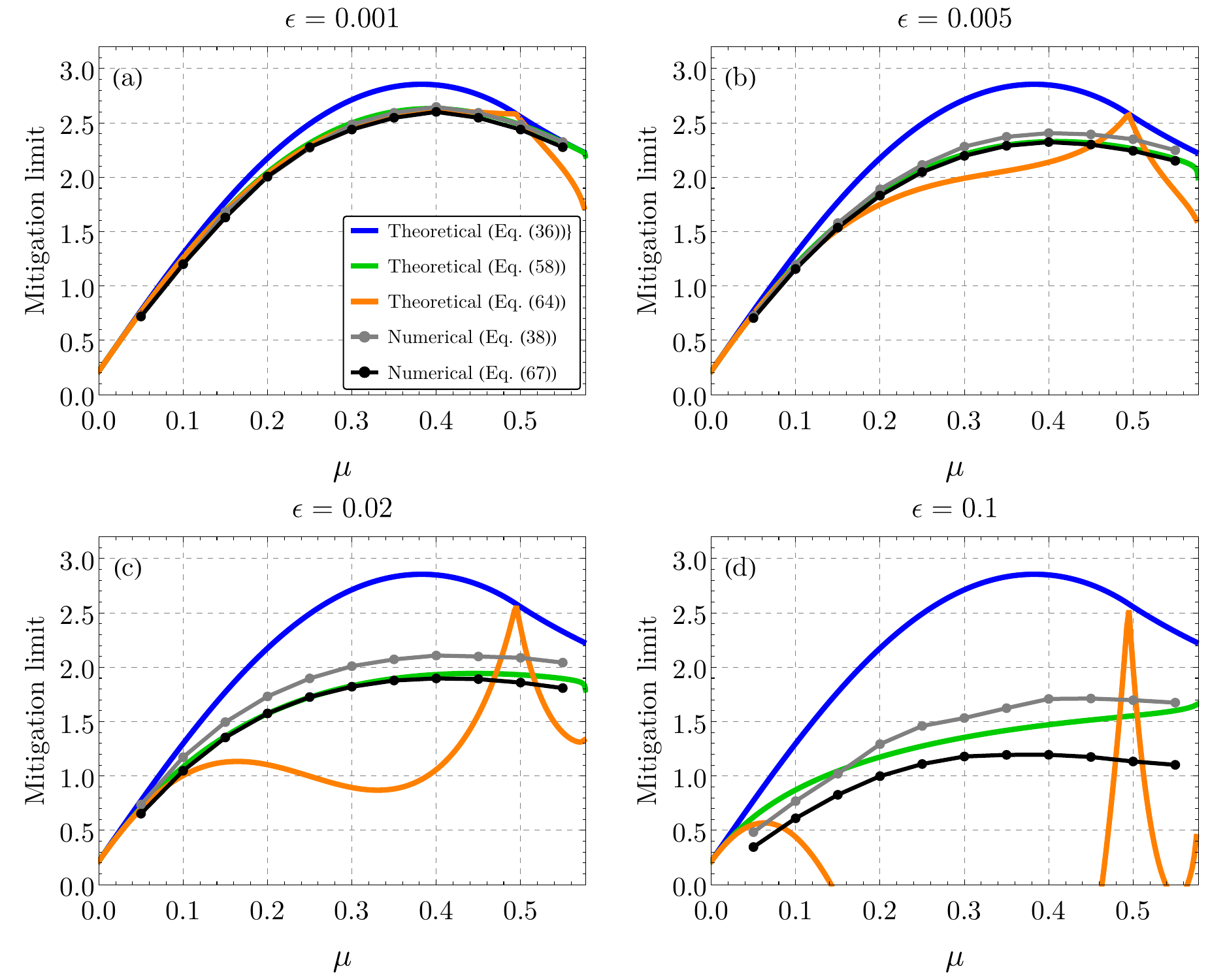}
	\caption{Theoretical mitigation limits: the one obtained within the zeroth-order approximation \eqref{eq:laidNES} (blue line) and those derived from the scaling law (solving numerically \cref{defNewrhoml} (green line) and the one given by \cref{eq:mitlimEps} (orange line)) compared with mitigation limits measured on direct numerical integration of the approximated slow flow \eqref{eq:slowRealSlow2} (gray dots) and on the full order system~\eqref{eq:airfiolnodim} (black dots). Parameters \eqref{eq:airfoilpar} and \eqref{eq:NESpar} are used and ((a) $\epsilon=0.001$, (b), $\epsilon=0.005$, $\epsilon=0.02$ and (d) $\epsilon=0.1$.}
	\label{fig:MitigLim}
\end{figure*}

\begin{figure*}[t!]
	\centering
	\includegraphics[width=0.9\columnwidth]{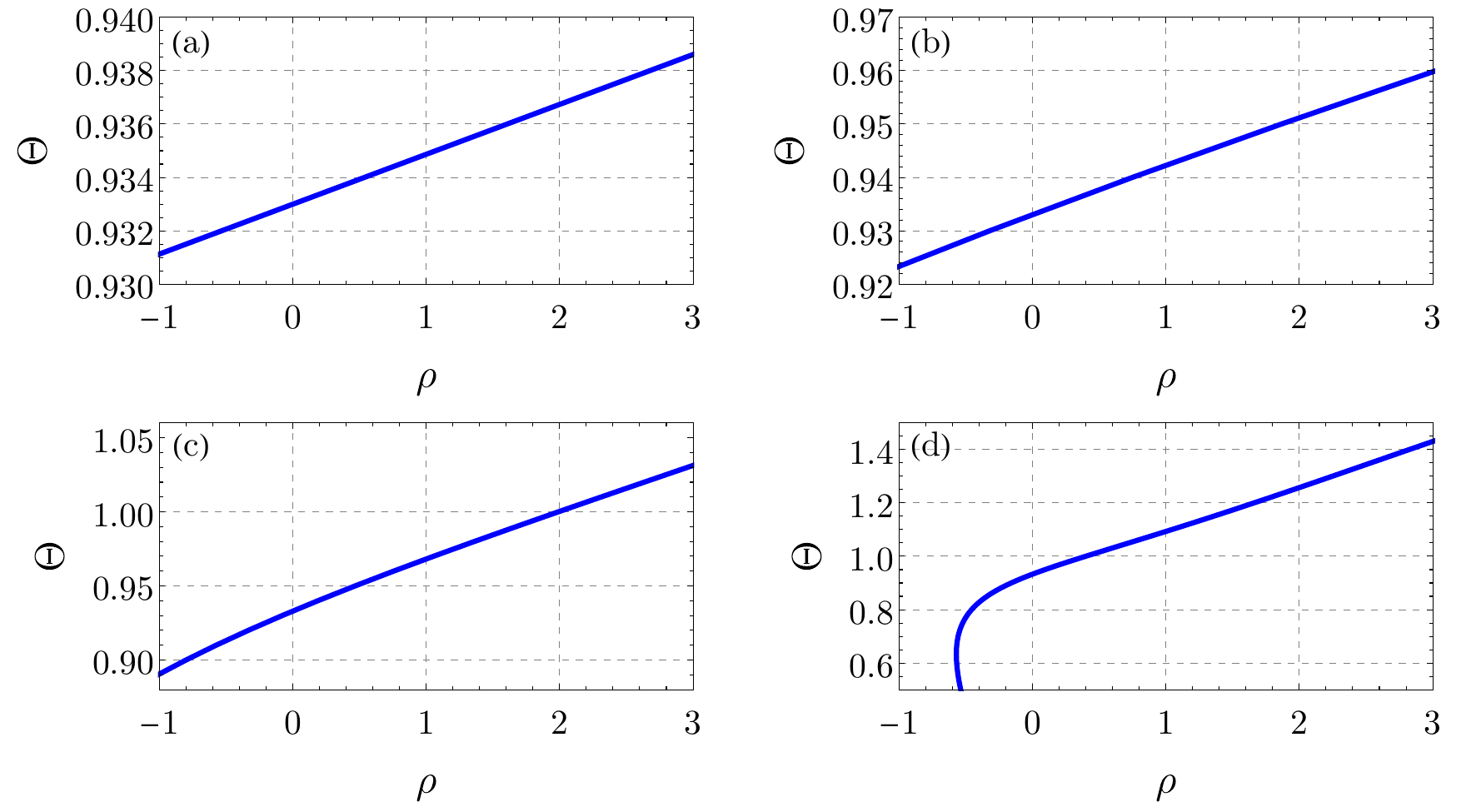}
	\caption{The physical bifurcation parameter $\Theta$ with respect to the generalized bifurcation parameter $\rho$. As in \cref{fig:MitigLim}, parameters \eqref{eq:airfoilpar} are used and (a) $\epsilon=0.001$, (b), $\epsilon=0.005$, $\epsilon=0.02$ and (d) $\epsilon=0.1$.}
	\label{fig:prho}
\end{figure*}

In \cref{fig:MitigLim}, the comparison in now performed in terms of mitigation limit. The theoretical predictions, the one obtained within the zeroth-order approximation \eqref{eq:laidNES} and those derived from the scaling law (solving numerically \cref{defNewrhoml} and the one given by \cref{eq:mitlimEps}) are compared with mitigation limits measured on direct numerical integration of the approximated slow flow \eqref{eq:slowRealSlow2} and on the full order system~\eqref{eq:airfiolnodim}. In these last two cases, direct numerical integrations are performed for increasing values of $\rho$  (by increasing $\Theta$) and the mitigation limit corresponds to the last value of $\rho$ allowing a harmless situation. In general one observes at this value the transition from a SMR to a ``no mitigation regime" as shown for example in \cref{fig:pFlutterExSimu}. All the mitigations limits are plotted with respect to the NES damping coefficient $\mu$ (with $\mu \in [0,1/\sqrt{3}]$\footnote{Remembering that in this work only situations in which \cref{eq:rocond} holds are considered.}) and for four values of the perturbation parameter $\epsilon$. One takes first a very little value ($\epsilon=0.001$) to be sure to respect the assumptions of the asymptotic analysis (i.e. $0<\epsilon\ll1$) and thus validate the method under these conditions. Then, three larger values are taken ($\epsilon=0.005$, 0.02 and 0.1) to evaluate the robustness of the method when one deviates from the assumption of a small $\epsilon$. To relate these results to the physical bifurcation parameter $\Theta$, the latter is plotted with respect to the generalized bifurcation parameter $\rho$ in \cref{fig:prho} for the same values of $\epsilon$ as in \cref{fig:MitigLim}.

First, in \cref{fig:MitigLim}, one can see that the zeroth-order approximation gives an inaccurate prediction, even for the smallest value of $\epsilon =0.001$. Moreover, one observes that, on the contrary, solving \cref{defNewrhoml} provides a very good prediction of the mitigation limits measured on numerical simulations. For the two following values of $\epsilon$ (0.005 and 0.02), these theoretical prediction is closer to the mitigation limits obtained from numerical simulations of the full order system~\eqref{eq:airfiolnodim} than to those measured on numerical integrations of the approximated slow flow \eqref{eq:slowRealSlow2}. This is surprising because \cref{defNewrhoml} is derived from \cref{eq:slowRealSlow2} but, in general, in these kind of unstable systems coupled to NES, the mitigation limits measured on the full order system are slightly smaller than those measured on the slow flow. Here, when $\epsilon$ increases the prediction depreciates becoming smaller than mitigation limits obtained from numerical simulations of \cref{eq:slowRealSlow2} and getting closer to those measured on full order system~\eqref{eq:airfiolnodim}. For the largest value $\epsilon=0.1$, the theoretical mitigation limit obtained solving \cref{defNewrhoml} (in green on the figure) depreciates again but it is still satisfactory in a qualitative point of view. Moreover one can see that the theoretical prediction (in green) no longer has a local maximum and the optimal value of $\mu$ is equal to $1/\sqrt{3}$, its limit value for the existence of the SMR regimes (see \cref{eq:rocond}).

Finally, the prediction provided by \cref{eq:mitlimEps} deteriorates enormously for large values of $\epsilon$. This may be due to the non-convergent nature of the series~\eqref{eq:powerser}, as already mentioned above (see comments after \cref{eq:mitlimEps}).

\begin{figure}[t!]%\sidecaption
	\centering
	\includegraphics[width=0.55\columnwidth]{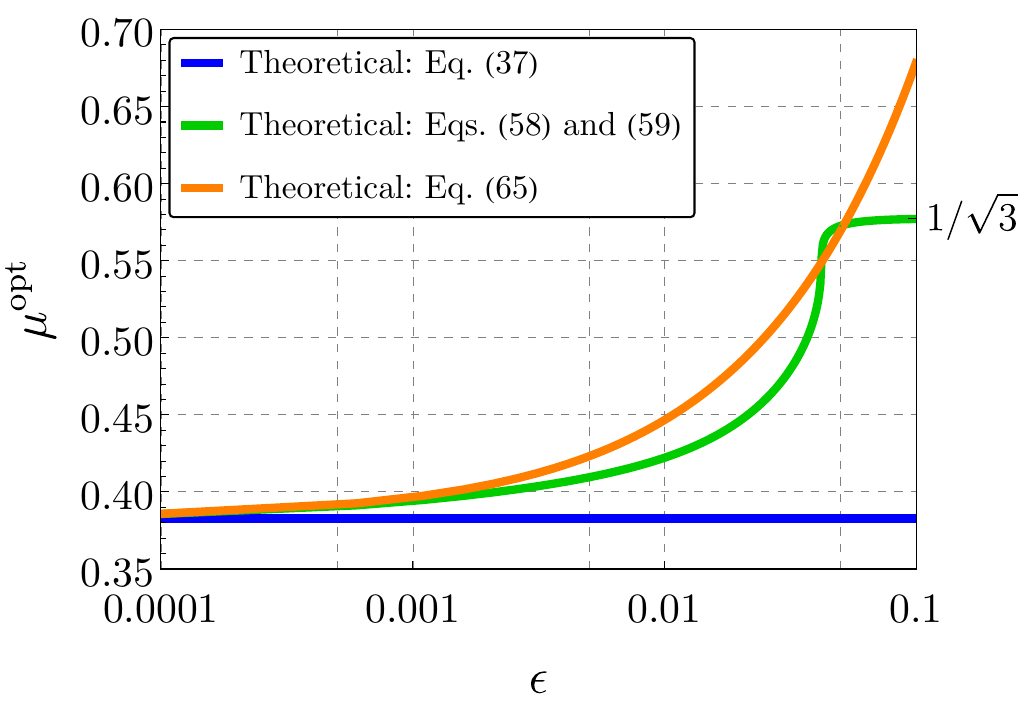}
	\caption{The theoretical optimal value of the NES damping coefficient with respect to the small perturbation parameter $\epsilon$. The one obtained from the zeroth-order approximation $\mu_0^\text{opt}$ given by \cref{eq:muopt0} (blue line) is compared to those derived from the scaling law \eqref{eq:srAi1}: the one obtained solving numerically $d_\mu \rho=0$ (with $d_\mu \rho$ given by \cref{eq:drho1}) together with \cref{defNewrhomlc} (green line) and the asymptotic expression given by \cref{eq:muoptEpsAs} (orange line). As previously, parameters \eqref{eq:airfoilpar} and \eqref{eq:NESpar} are used.}
	\label{fig:pmuOpt}
\end{figure}

Finally, in \cref{fig:pmuOpt}, the theoretical optimal value of the NES damping coefficient obtained from the zeroth-order approximation $\mu_0^\text{opt}$ (see \cref{eq:muopt0}) is compared to those derived from the scaling law \eqref{eq:srAi1}: the one obtained solving numerically $d_\mu \rho=0$ (with $d_\mu \rho$ given by \cref{eq:drho1}) together with \cref{defNewrhomlc} and the asymptotic expression given by \cref{eq:muoptEpsAs}. All this is plotted with respect to the perturbation parameter~$\epsilon$. Assuming a good agreement between the theoretical prediction of the mitigation limit obtained solving \cref{defNewrhoml} and the actual value (as shown by \cref{fig:MitigLim}), only theoretical values are presented in \cref{fig:pmuOpt}. The figure highlights the relevance of the proposed results especially for large values of the parameter $\epsilon$. We can also see that the asymptotic expression is here able to describe qualitatively the evolution of $\mu^\text{opt}$ with respect to the perturbation parameter $\epsilon$ for $\epsilon \lesssim 0.04$. For $\epsilon \gtrsim 0.04$, the prediction obtained from \cref{eq:drho1,defNewrhomlc} tends to $1/\sqrt{3}$ whereas the prediction given by \cref{eq:muoptEpsAs} continues to grow.

%-------------------------------------------------------------------------------------------------%
%-------------------------------------------------------------------------------------------------%
% Section
%-------------------------------------------------------------------------------------------------%
%-------------------------------------------------------------------------------------------------%
\section{Conclusion}
\label{sec:ccl}

In this work a mechanical system with one unstable mode coupled to a Nonlinear Energy Sink (NES) has been studied. The equations of motion of the system have been first simplified using a reduced-order model for the primary structure by keeping only its unstable modal coordinates. Introducing a small perturbation parameter related to the mass ratio between the NES and the primary structure, the slow flow has been then derived by means of the complexification-averaging method. Because of the presence of the small perturbation parameter, the slow flow is governed by two different time scales. More precisely, within the framework of the geometric singular perturbation, in its real form it appears as a $(2, 1)$-fast-slow system. The critical manifold of this slow flow has been computed. Within the zeroth-order approximation, i.e. the analysis within the limit case in which the perturbation parameter is equal to zero, the solution of the slow flow is approximated by a discontinuous phase trajectory in which the fast and slow epochs are described independently of each other. In particular, the slow epoch are supposed to occur on attracting parts of the critical manifold. Moreover, in this context a theoretical prediction of the mitigation has been obtained. The mitigation limit being defined as the value of the generalized bifurcation parameter, which is obtained rescaling the eigenvalue real part of the primary structure corresponding to the unstable mode, which separates harmless situations (in which the NES acts) from the harmful situation (in which the NES does not act). The results obtained within the zeroth-order approximation (both phase trajectory and mitigation limit) depreciate for the largest values of the perturbation parameter (the latter remaining small in the context of this work).

The novelty of the present work is to use the center manifold theorem to reduce the slow flow, in the neighborhood of a fold point of its critical manifold, to a normal form of the dynamic saddle-node bifurcation. This allowed us to obtain a scaling law for the slow flow in this neighborhood, i.e.  a law which describes the dependance, with respect to the small perturbation parameter, of the distance between the critical manifold and the actual trajectory of the slow flow. The law reveals that the slow flow scales in a nontrivial way with respect to the perturbation parameter, involving the fractional exponents 1/3 and 2/3. Finally, a theoretical prediction of the mitigation limit has been deduced from the scaling law. With certain assumptions, it was shown that this prediction is a correction with respect to the prediction obtained within the zeroth-order approximations, involving also a fractional order dependence with respect to the perturbation parameter.

The last section of the paper has been devoted to illustrate and validate the proposed methodology on a an aeroelastic aircraft wing model coupled to one NES. The accuracy of the present theoretical results compared to those obtained on numerical simulations of the full order system together with the general nature of the method suggest that the latter may constitute a tool for the practical design of NES.

%-------------------------------------------------------------------------------------------------%
% Ackowledgments
%-------------------------------------------------------------------------------------------------%

\section*{Acknowledgments}

We thank the anonymous reviewers for their careful reading of our manuscript and their many relevant comments and suggestions.

%-------------------------------------------------------------------------------------------------%
% Bibliographie
%-------------------------------------------------------------------------------------------------%
\bibliographystyle{elsarticle-num}\small
\bibliography{Biblio_NES_BBergeot}

\end{document}